\documentclass[twocolumn,preprintnumbers,amsmath,amssymb]{revtex4-1}
\usepackage{colortbl}
\usepackage[table]{xcolor}
\usepackage{array}
\usepackage{booktabs}
\usepackage{subfigure}
\usepackage{graphicx}
\usepackage{dcolumn}
\usepackage{bm}

\usepackage{color}
\definecolor{darkgreen}{rgb}{0.1,.6,.1}
\definecolor{greenblue}{rgb}{0.0,.1,.4}

\usepackage{array}
\usepackage{makecell}

\usepackage{tikz}

\definecolor{half}{rgb}{0.91, 0.84, 0.42}
\definecolor{U50to59}{rgb}{0.47,0.8,0.47}
\definecolor{U60to100}{rgb}{0.8,0.15,0.15}
\definecolor{L50to59}{rgb}{0.94,0.5,0.5}
\definecolor{L60to100}{rgb}{0.15,0.55,0.15}

\usepackage{xcolor,colortbl}

\usepackage{ragged2e}

\usepackage[normalem]{ulem}

\begin{document}



\title{Regime shifts driven by dynamic correlations in gene expression noise}

\author{Yogita Sharma} 
\affiliation{Department of Mathematics, Indian Institute of Technology
  Ropar, Punjab 140 001, India}

\author{Partha Sharathi Dutta} 
\thanks{Corresponding author: parthasharathi@iitrpr.ac.in} 
\affiliation{Department of Mathematics, Indian Institute of Technology
  Ropar, Punjab 140 001, India}

\received{:to be included by reviewer}
\date{\today}

\begin{abstract}

Gene expression is a noisy process that leads to regime shift between
alternative steady states among individual living cells, inducing
phenotypic variability.  The effects of white noise on the regime
shift in bistable systems have been well characterized, however little
is known about such effects of colored noise (noise with non-zero
correlation time).  Here, we show that noise correlation time, by
considering a genetic circuit of autoactivation, can have significant
effect on the regime shift in gene expression.  We demonstrate this
theoretically, using stochastic potential, stationary probability
density function and first-passage time based on the Fokker-Planck
description, where the Ornstein-Uhlenbeck process is used to model
colored noise.  We find that increase in noise correlation time in
degradation rate can induce a regime shift from low to high protein
concentration state and enhance the bistable regime, while increase in
noise correlation time in basal rate retain the bimodal distribution.
We then show how cross-correlated colored noises in basal and
degradation rates can induce regime shifts from low to high protein
concentration state, but reduce the bistable regime.  In addition, we
show that early warning indicators can also be used to predict shifts
between distinct phenotypic states in gene expression.  Predictions
that a cell is about to shift to a harmful phenotype could improve
early therapeutic intervention in complex human diseases.

\end{abstract}

\maketitle

\section{Introduction}

Natural systems can undergo sudden, large and irreversible changes
under the influence of small stochastic perturbations
\cite{Book_Sch,ScJo09}. Such qualitative sudden changes are known as
``regime shifts'' have been found in a variety of ecological systems
\cite{Sc01,ScCa2003,Wang:2012}, climate systems \cite{LeHeKr08},
biological systems \cite{Ve05,Mc03,kor2014,Ri2016}, financial markets
\cite{MaLeSu08}, physical systems \cite{FoWh15,GoSh:2016}, etc.  It
has been identified that regime shifts generally occur at tipping
points (namely bifurcation points) \cite{ScJo09,Scheffer:2012sc},
where the system abruptly shifts from one stable state to another
stable state.  There are also examples of purely noise
induced regime shifts (known as stochastic switching)
\cite{DaCa15,SKP14,Sh2016}.  Regime shifts have the potential to
invoke serious and harmful consequences for environment as well as
human well-being \cite{Sc01,FoWh15,Ri2016}.

Understanding the mechanisms of regime shifts and predicting them
using early warning signals (EWS) have been recently emerged as a
challenging area of research due to the potential application in
management and prevention of sudden catastrophes in complex systems.
Numerous studies have been carried out to develop EWS for successfully
predicting regime shifts
\cite{Dakos:2012pone,Scheffer:2012sc,CaBr08,GuJa08,ScJo09}.  Extensive
research on EWS suggests that statistical signatures, such as
concurrent increase in ``variance'', ``autocorrelation'', ``skewness''
can predict regime shifts in a wide variety of complex systems
\cite{ScJo09,Scheffer:2012sc,GuJa08}.  These EWS are mainly derived
from the phenomenon of critical slowing down, which is associated with
a tipping point at which the stability of an equilibrium state changes
as the dominant real eigenvalue becomes zero
\cite{Book_Sch,ScJo09,Scheffer:2012sc}.  As a result, rate of recovery
from small stochastic perturbations becomes slow as the system
approaches a tipping point, resulting concurrent increase in variance,
autocorrelation and skewness prior to a regime shift.  However,
sometimes these EWS are not present before a regime shift due to
statistical limitations and confer false alarms \cite{sc2015}.  In
such a situation, apart from the aforementioned indicators, other
indicators, e.g., ``conditional heteroskedasticity" \cite{Se11,En1982}
can be very useful to detect regime shifts.  Conditional
heteroskedasticity is used to investigate the possible links between
time series data and their volatilities \cite{En1982}.  This indicator
generally avoids the chance of false alarms as it is associated with
significant test and their probabilities.  Majority of the earlier
studies on predicting regime shifts using EWS have focused on the
ecological and climate systems \cite{Book_Sch,DaCa15,sc2015}.
However, few recent studies have reported the huge potential of EWS as
risk markers from molecular biology to chronic human diseases
\cite{PaPaBo13,kor2014,gl15,TrAn15,Sh2016,Ri2016}.

Regime shifts those arise in medical conditions can increase the risk
of diseases and even result in sudden death \cite{gl15,Ri2016}.
Recently, EWS for detecting regime shifts in ecology have got special
attention in medical sciences \cite{ChLi12,kor2014,Ri2016}. The
ability to predict such regime shifts could prove fruitful in early
detection of diseases \cite{Van2014,La2012,kra2012,Sc2013}. An
important example of regime shift in molecular biology is genetic
regulatory system, which includes sudden transition in protein
production level in individual cells resulting disease onset
\cite{Sm98}. In genetically identical cells fluctuations in
transcription and translation give rise to regime shifts between
alternative states (i.e., phenotypic variability) in intracellular
protein concentrations \cite{KaEl05}.  Indeed, in positive-feedback
regulation individual cells can exist in different steady states, some
live in the ``on'' expression state and others live in the ``off''
expression state \cite{Ha00}.  These ``on'' and ``off'' states are
mainly related with protein production.  Also, the cells perform a
range of specialized functions for protein production that depend upon
gene expression states.  For instance, $\beta$ cells in the pancreas
produce the protein hormone insulin that depends upon HLA-encoding
gene states, $\alpha$ cells produce the hormone glucagon, lymphocytes
of the immune system produce antibodies-proteins (gamma globulin's),
while developing red blood cells produce the oxygen-transport protein
hemoglobin.  Finding the causes of regime shift and predicting that a
cell is about to shift to a harmful gene expression state can improve
critical care management for complex human diseases.

In previous studies, the stochastic fluctuations associated with gene
expression are considered as Gaussian white noise (noise with zero
correlation time) \cite{Ha00,LiJi04,Ch08,Frcasaib12,Gh12}.  In contrast,
few recent studies have shown that gene expression noise can also be
colored in nature (noise with non-zero correlation time)
\cite{Ro2005,Si2006,ShOSw08,du2008,Dunlop:thesis2008}.  These studies
have measured the variability of protein levels in human cellular
system and showed that cell to cell variability of protein levels can
be correlated over generations \cite{Si2006}. It has also been
measured experimentally that gene expression noise has a finite
correlation time \cite{ShOSw08}.  Moreover, colored noise can break
bistability in different ways than that of white noise \cite{Kl1988}.
In order to address these issues, it is important to study the effects
of noise correlation on regime shifts in gene expression.

One of the key questions addressed in this paper is: How the dynamic
correlation in gene expression noise affects the characteristics of
sudden regime shifts between alternative steady states (i.e., low and
high protein concentration states)?  For this, we begin with a
stochastic version of gene regulatory system: a genetic autoactivating
switch.  The colored noise is modeled using Ornstein-Uhlenbeck
process.  We compute the stochastic potential and the stationary
probability density function to quantify the effects of noise
intensity and correlation time on the relative stability of
alternative steady states using Fokker-Planck description. We then
obtain the mean first-passage time (MFPT) for escape over the
potential barrier.  We show that increase in the noise correlation
time in degradation rate can induce a regime shift from low to high
protein concentration state and enhance the bistable regime, while
noise in basal rate retain the bimodal distribution of the system
steady states.  We also show that cross-correlated colored noises in
basal and degradation rates can induce regime shifts from low to high
protein concentration state, however reduce the bistable regime.
Further, we examine EWS prior to a regime shift in gene expression
dynamics, which can prove to be very useful to predict that a cell is
about to shift to a harmful phenotype.

The paper is organized as follows: Section~\ref{sec:2} presents the
description of a stochastic model of gene expression. In
Sec.~\ref{sec:3a}, steady state analysis of the stochastic model is
presented.  Impacts of noise correlation time, noise intensity and
cross-correlation strength on the effective potential landscape and
the stationary probability density function are calculated in
Sec.~\ref{sec:3b}.  We then examine the MFPT of the system driven by
the correlated noise in Sec.~\ref{sec:3c}, and precursors of regime
shift in Sec.~\ref{sec:3d}.  Finally, in Sec.~\ref{sec:dis}, we
conclude the study by discussing the key findings reported in this
paper.

\begin{figure}
\begin{center}
\includegraphics[width=0.95\columnwidth]{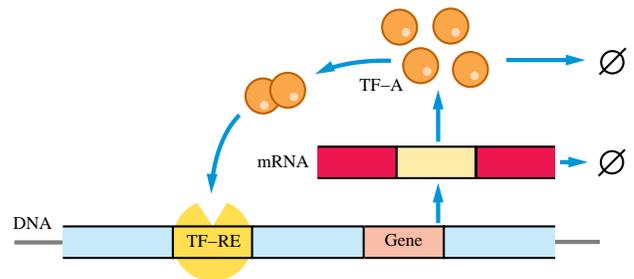}
\caption{\label{sch} (Color online) A schematic of the autoactivating
  genetic switch.  The expression of gene leads to protein monomers
  (TF-A) and after oligomerization they bind to the upstream
  regulatory site (TF-RE), activating production of monomers.
  Degradation of protein and mRNA are denoted by the slashed circles
  ($\varnothing$).}
\end{center}
\end{figure}


\section{A stochastic model of gene expression}\label{sec:2}

To understand the effects of noise correlation, we consider a well
studied stochastic model of gene expression: the autoactivating
switch, which consists multiple stable states
\cite{BeSeSe01,TyChNo03,Ch08,Frcasaib12}.  A schematic picture of the
genetic circuit is shown in Fig.~\ref{sch}, which involves a single
gene that transcribes a single protein called activator TF-A.  On
dimerization the protein TF-A dimer stimulates transcription when
binds to the responsive element TF-RE in the DNA sequence.  The mRNA
produced in transcription and protein monomer produced in translation
then follow post-transcriptional degradation which is an important
regulatory step (see Fig.~\ref{sch}) \cite{Is2003}.  Letting $x(t)$
and $y(t)$ as concentrations of the activator protein TF-A and the
mRNA respectively, we can write the rate equations describing the
evolution of $x(t)$ and $y(t)$:
\begin{subequations}\label{eq1}
\begin{align}
\frac{dx}{dt} & = Ky-k_{deg_{r}} x,\\ \frac{dy}{dt} & =
F(x)-k_{deg_{m}} y,
\end{align}
\end{subequations}
where the parameter $K$ is the translation rate, $F(x)$ is the mRNA
transcription rate, $k_{deg_{r}}$ and $k_{deg_{m}}$ are the
degradation rates of the protein monomers and the mRNA. The function
$F(x)$ is given by a Hill-type function \cite{Tha2001}:
$$F(x)=\frac{k_{max}\:x^{\mathcal H}}{k_{d}+x^{\mathcal H}}+k_f,$$ where
  $k_{max}$ is the maximum transcription rate, $k_{d}$ is the Hill
  constant, $k_{f}$ is the basal transcription rate and $\mathcal H$
  is the Hill coefficient which we consider $\mathcal H=2$
  \cite{ZhYaTa11}.  The degradation rate of mRNA molecules is usually
  much faster than that of proteins \cite{Tha2001}, i.e.,
  $k_{deg_{m}}>> k_{deg_{r}}$.

\begin{figure}[!h]
\begin{center}
\includegraphics[width=0.8\columnwidth]{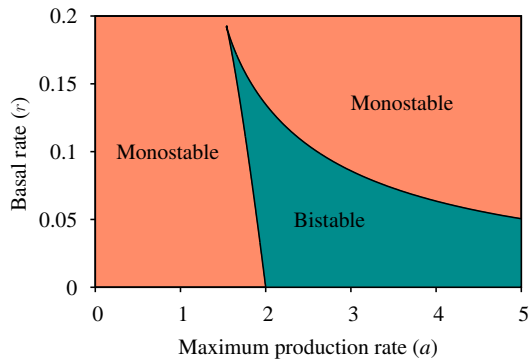}
\caption{\label{sch11} (Color online) Phase diagram of the gene
  expression model (\ref{Eq:Fl}) in $(a,r)-$plane. The curve
  separating the monostable region from the bistable region is a
  saddle-node bifurcation curve.}
\end{center}
\end{figure}

Since the fast reactions equilibrate quickly, to reduce the dimension
of the system it is useful to apply the quasi-steady state
approximation (QSSA) which replaces state variables involved in the
fast reactions with their equilibrium values. This dimension reduction
greatly simplifies the complexity of the system \cite{Ra03}.  Now
employing the QSSA in Eq.~(\ref{eq1}) by replacing the equilibrium
value $y=F(x)/k_{deg_m}$ from the ``fast'' Eq.~(\ref{eq1}b) into the
``slow'' Eq.~(\ref{eq1}a) and taking $\mathcal H=2$, we obtain the
following  reduced system:
\begin{equation}\label{eq2}
\frac{dx}{dt}=\frac{K}{k_{deg_m}}\left(\frac{k_{max}x^{2}}{k_{d}+x^2}
+k_{f}\right)-k_{deg_r} x \:.
\end{equation} 
The above Eq.~(\ref{eq2}) can also be written as \cite{WeBu13}:
\begin{equation}\label{eq3}
\frac{dx}{dt} = R+a\frac{x^2}{k_{d}+x^2}-k_{deg_{r}}x,
\end{equation}
where $R=\frac{k_{f}K}{k_{deg_{m}}}$ is the basal expression rate and
$a=\frac{k_{max}K}{k_{deg_{m}}}$ is the maximum transcription rate.
Now the dimensionless version of equation Eq.~(\ref{eq3}) is:
\begin{equation}\label{Eq:diml}
\frac{d\tilde{x}}{d\tilde{t}}=\tilde{r}+\tilde{a} \frac{\tilde{x}^2}
     {1+\tilde{x}^2} -\tilde{x},
\end{equation}
where $\tilde{x}= \frac{x}{\sqrt{k_{d}}}$, $\tilde{t}=k_{deg_{r}}t$,
$\tilde{a}=\frac{a}{k_{deg_{r}}\sqrt{k_{d}}}$, and
$\tilde{r}=\frac{R}{k_{deg_{r}}\sqrt{k_{d}}}$.  Finally, we use $x$,
$t$, $r$ and $a$ in place of $\tilde x$, $\tilde t$, $\tilde r$ and
$\tilde a$, and Eq.~(\ref{Eq:diml}) reads:
\begin{equation}\label{Eq:Fl}
\frac{dx}{dt}=r+a \frac{x^2}
{1+x^2} -x\;.
\end{equation}
For a range of $a$, if $0 < r < 1/3\sqrt{3} \approx 0.19245\:$ then
Eq.~(\ref{Eq:Fl}) exhibits two types of asymptotic behaviors:
monostability and bistability (i.e., it leads to phenotypic
variability) \cite{WeBu13}. In the case of bistability the system has
three equilibrium points, the middle one (say $x_{u}$) is unstable and
the other two are stable. In the bistable regime, the initial
condition (say $x_{i}$) plays a key role in determining the final
equilibrium state of the system.  All the initial values $x_i>x_{u}$
will evolve to the upper equilibrium point and others $x_i<x_{u}$ will
evolve to the lower equilibrium point in the stationary state.
Figure~\ref{sch11} depicts the phase diagram of the
model~(\ref{Eq:Fl}) in the $(a,r)$-plane for different values of the
control parameters $a$ and $r$.  The region of bistability is bounded
by a saddle-node bifurcation curve at which transition occurs from
monostable to bistable regime or vice versa.  A thorough analysis of
the deterministic model~(\ref{Eq:Fl}) is given in \cite{Sm98}.

As already discussed in the introduction, here we are mainly
interested in understanding the effects of correlated gene expression
noise on the regime switching between two alternative steady states.
Therefore, in the model (\ref{Eq:Fl}) we incorporate {\em correlated}
stochastic process in the form of two fluctuating rates.  We assume
that variability in the basal and the degradation rates causes the
production rate of protein to fluctuate \cite{LiJi04}.  That is, in
Eq.~(\ref{Eq:Fl}) the basal rate varies stochastically as $r
\rightarrow r+\eta(t)$ and also the degradation rate varies
stochastically as $1 \rightarrow 1+\xi(t)$.  We consider $\xi(t)$ and
$\eta(t)$ to be Ornstein-Uhlenbeck (OU) processes \cite{Ga85}:
positively correlated Gaussian noise (i.e., {\em colored Gaussian
  noise}) with a zero mean and correlation time $\tau_1$ and $\tau_2$,
respectively.  The Langevin equation corresponding to
Eq.~(\ref{Eq:Fl}) which contains both the Gaussian colored noises
$\eta(t)$ and $\xi(t)$ can be written as \cite{Ha00,LiJi04}:
\begin{eqnarray}
\frac{dx}{dt}&=&(r+\eta(t)) + a\frac{x^2}{1+x^2}- (1+\xi(t))x,
\nonumber\\
&=&f(x)+g_{1}(x) \xi(t)+g_{2}(x) \eta(t), \label{eq:stc}
\end{eqnarray}
where $f(x)=r+ \frac{ax^2}{1+x^2}-x$, $g_{1}(x)=-x$ and $g_{2}(x)=1$.
Thus, here the noise $\xi(t)$ can be considered as multiplicative
colored noise in comparison to $\eta(t)$, which works as additive
colored noise \cite{Ha00}.  The OU processes $\xi(t)$ and $\eta(t)$
satisfy the following equations:
\begin{eqnarray}
\dot{\xi}(t) & = & -\frac{\xi(t)}{\tau_{1}} + \frac{\sqrt{2\sigma_1}}{\tau_1}\xi_{1}(t),
\nonumber\\ 
\dot{\eta}(t) & = & -\frac{\eta(t)}{\tau_{2}} + \frac{\sqrt{2\sigma_2}}{\tau_{2}}
\eta_{1}(t),\nonumber 
\end{eqnarray}
where $\xi_{1}(t)$ and $\eta_{1}(t)$ are white Gaussian noises with
zero mean and unit variance \cite{Ga85}.  The parameters $\sigma_{i}$
and $\tau_{i}(\neq 0)$, for $i=1,2$ are noise strength and self
correlation time of $\xi(t)$ and $\eta(t)$, respectively.  The colored
Gaussian noises $\xi(t)$ and $\eta(t)$ satisfy the following
statistical properties:
\begin{subequations}\label{Eq:pro}
\begin{align}
& \langle \xi(t) \rangle = \langle \eta(t) \rangle = 0,\nonumber \\
& \langle \xi(t)\xi(t') \rangle =
 (\sigma_{1}/\tau_{1})\exp(-|t-t'|/\tau_{1}),\nonumber \\
& \langle \eta(t)\eta(t') \rangle = 
(\sigma_{2}/\tau_{2})\exp(-|t-t'|/\tau_{2}),\nonumber \\
& \langle \xi(t)\eta(t') \rangle = 
(\lambda \sqrt{\sigma_{1}\sigma_{2}}/\tau_{3})\exp(-|t-t'|/\tau_{3}),\nonumber
\end{align}
\end{subequations}
where $\lambda$ measures the coupling strength between $\xi(t)$ and
$\eta(t)$, $\tau_{3}$ is the correlation time between the noises,
while $t$ and $t'$ denote two different moments.

In order to understand the influence of colored noises on the rapid
switching between two alternative stable states, we employ theoretical
calculations of probability densities, potential functions, and MFPTs
of Eq.~(\ref{eq:stc}).


\section{Results}

\subsection{Steady state analysis of the stochastic system}\label{sec:3a}

To solve the stochastic Eq.~(\ref{eq:stc}), we begin with the
probability density $P(x,t)$, which is the probability that the
protein concentration will attain the value $x$ at time $t$.  The
approximate Fokker-Planck equation (AFPE) for $P(x,t)$ corresponding
to Eq.~(\ref{eq:stc}) is \cite{Sa1982,liang2004}:
\begin{equation}\label{eq:ps}
\frac{\partial P(x,t)}{\partial t}  =  -\frac{\partial}{\partial x}
\left[A(x) P(x,t)\right]+\frac{\partial^2}{\partial^2 x}[B(x) P(x,t)],
\end{equation}
where,
\begin{subequations}\label{eq:exp1}
\begin{align}
A(x) & = f(x)+\frac{\sigma_{1}x}{1-\tau_{1}f'(x_{s})}-\frac{\lambda
  \sqrt{\sigma_{1}\sigma_{2}}}{1-\tau_{3}f'(x_{s})},  \\ 
B(x) & = \frac{\sigma_{1}x^2}{1-\tau_{1}f'(x_{s})}-\frac{2\lambda
  \sqrt{\sigma_{1}\sigma_{2}}x}{1-\tau_{3}f'(x_{s})}
+\frac{\sigma_{2}}{1-\tau_{2}f'(x_{s})}, 
\end{align}
\end{subequations}
and $f'(x_{s})$ is the derivative of $f(x)$ at the equilibrium
point $x_{s}$.  The derivative $f'(x_{s})$ is given by:
\begin{equation*}
f'(x_{s})=\frac{2ax_{s}}{(1+x_{s}^2)^2}-1, 
\end{equation*}
where the equilibrium point $x_{s}$ is:
\begin{eqnarray}
x_{s} & = &
\sqrt[3]{-\frac{m}{2}+\sqrt{\Big(\frac{m}{2}\Big)^2
+\Big(\frac{l}{3}\Big)^3}}\nonumber\\ &&
+\sqrt[3]{-\frac{m}{2}-\sqrt{\Big(\frac{m}{2}\Big)^2
+\Big(\frac{l}{3}\Big)^3}}-\frac{n}{3},
\end{eqnarray}
with $l$, $m$ and $n$ are as: $l = 1-\frac{(r+a)^2}{3}$, $m =
\frac{1}{27}(r+a)^3+\frac{l}{3}(r+a)-r$, and $n = -(r+a)$.  The point
$x_{s}$ is the only real solution of $f(x)=0$.  Relation between the
two functions $A(x)$ and $B(x)$ are given by:
\begin{equation*}
A(x)=f(x)+\frac{1}{2}\frac{d}{dx}B(x) .
\end{equation*} 
Moreover, the AFPE (\ref{eq:ps}) is valid for $1-\tau_{i}f'(x_{s})>0$
($i=1,2,3$) \cite{liang2004}.

The stationary probability density function (SPDF) $P_s(x)$ of $x$,
which is the stationary solution of the AFPE~(\ref{eq:ps}), is given
by:
\begin{eqnarray}
P_{s}(x) & = & \frac{N_{c}}{B(x)} \exp \Big[\int^{x}\frac{A(u)}{B(u)} du\Big]\nonumber\\
 & = & \frac{N_{c}}{\frac{\sigma_{1}x^2}{1-\tau_{1}f'(x_{s})}-\frac{2\lambda\sqrt{\sigma_{1}\sigma_{2}}x}{1-\tau_{3}f'(x_{s})}+\frac{\sigma_{2}}{1-\tau_{2}f'(x_{s})}}\times \nonumber\\
&& \!\!\!\!\!\!\!\!\!
\exp\left[\int^x\frac{f(u)+\frac{\sigma_{1}u}{1-\tau_{1}f'(x_{s})}-\frac{\lambda\sqrt{\sigma_{1}\sigma_{2}}}{1-\tau_{3}f'(x_{s})}}
{\frac{\sigma_{1}u^2}{1-\tau_{1}f'(x_{s})}-\frac{2\lambda\sqrt{\sigma_{1}\sigma_{2}}u}{1-\tau_{3}f'(x_{s})}+\frac{\sigma_{2}}{1-\tau_{2}f'(x_{s})}}du \right],\nonumber\\
\label{Eq:Spdf2}
\end{eqnarray}
where $N_{c}$ is normalization constant obtained from:
\begin{equation*}
\int_{0}^{\infty}P_{s}(x)dx=1.
\end{equation*}
In analogy with the physical situation of a particle moving in a
potential, the SPDF peaks correspond to the valleys of the potential
(i.e., attractors) and troughs correspond to the tops of the potential
(i.e., repellors).  We can also introduce a stochastic potential by
writing the SPDF (\ref{Eq:Spdf2}) in the form:
\begin{equation}
P_{s}(x)=N_{c}e^{-\phi(x)},
\end{equation}
where
\begin{eqnarray} 
\phi(x) && =  \frac{1}{2}\ln\left[ \frac{\sigma_{1}x^2}{1-\tau_{1}f'(x_{s})}-\frac{2\lambda\sqrt{\sigma_{1}\sigma_{2}}x}{1-\tau_{3}f'(x_{s})}+\frac{\sigma_{2}}{1-\tau_{2}f'(x_{s})}\right] \nonumber\\
&& -\int^{x}\frac{f(u)}{ \frac{\sigma_{1}u^2}{1-\tau_{1}f'(x_{s})}-\frac{2\lambda\sqrt{\sigma_{1}\sigma_{2}}u}{1-\tau_{3}f'(x_{s})}+\frac{\sigma_{2}}{1-\tau_{2}f'(x_{s})}}du,   \label{Eq:POT1}
\end{eqnarray}
is the stochastic potential of the system.  The stochastic potential
provides information about the the relative stability of the steady
states, likewise the deterministic potential of a system.

It is also important to know the stationary state of the system for
arbitrary noise intensities.  More specifically, we are interested in
understanding the transition phenomena between stationary states that
occur due to the presence of correlated noise.  For the deterministic
model (\ref{Eq:Fl}), this can be best visualized by the corresponding
bifurcation diagram representing the equilibrium protein concentration
$x$, for a range of control parameter.  In the stochastic model
(\ref{eq:stc}), a qualitative change in the stationary state is
accurately reflected by the behavior of the extrema of the SPDF
$P_s(x)$ \cite{WhoRle84}.  The extrema of $P_s(x)$ can easily be found
form the equation given below \cite{WhoRle84}:
\begin{equation}\label{Eq:Ext}
f(x)-\frac{\sigma_{1}x}{1-\tau_{1}f^{'}(x_{s})}
+\frac{\lambda\sqrt{\sigma_{1}\sigma_{2}}}{1-\tau_{3}f^{'}(x_{s})}=0.
\end{equation}

Using the above steady state calculations of the stochastic model
(\ref{eq:stc}), in next subsection we mainly focus on the dynamical
consequences due to the presence of dynamic correlations in noise.


\subsection{Effective potential landscape and stationary probability 
density function}\label{sec:3b}

In order to study the effects of variations in the stochastic
parameters (i.e., $\sigma_i$, $\tau_i$ and $\lambda$), we use the
evolution equation for SPDF (\ref{Eq:Spdf2}).  The SPDF, potential
function and extrema of SPDF are examined for three different cases:
(i) When noise is present only in the degradation rate.  (ii) When
noise is present only in the basal rate.  (iii) When noise is present
in both the rates, respectively.  In Table~I, we summarize the values
of stochastic parameters corresponding to the above three cases.

\begin{table}[!ht] \label{Tab1}
\label{t1}
\centering
\caption{Stochastic parameter values corresponding to three different
  cases: Colored noise in (i) the degradation rate, (ii) the basal
  rate and (iii) noise in both the degradation and basal rates.}
\begin{tabular}{l l l l l l l}
\hline\hline
\multicolumn{7}{c}{ \vspace{-.09in}}\\
\multicolumn{7}{c}{ $~$ Parameters:   \hspace{0.1in} $\sigma_{1}$ 
\hspace{0.22in} $\sigma_{2}$ \hspace{0.22in} $\lambda$ \hspace{0.22in}
$\tau_{1}$ \hspace{0.22in} $\tau_{2}$ \hspace{0.22in}
$\tau_{3}$}\\ \hline\hline
\vspace{-.1in} &&&&&&\\ $~~~$ {\em Case (i)}: $~$ & $\neq0$ & $~~~$
$=0$ & $~~~$ $=0$ & $~~~$ $\neq0$ & $~~~$ $=0$ & $~~$ $=0$\\
\vspace{-.12in}
 &&&&&&\\
\hline
\vspace{-.1in} &&&&&&\\ $~~~$ {\em Case (ii)}: $~$ & $=0$ & $~~~$
$\neq0$ & $~~~$ $=0$ & $~~~$ $=0$ & $~~~$ $\neq0$ & $~~$ $=0$\\
\vspace{-.12in}
&&&&&&\\
\hline
\vspace{-.1in} &&&&&&\\ $~~~$ {\em Case (iii)}: $~$ & $\neq0$ & $~~~$
$\neq0$ & $~~~$ $\neq0$ & $~~~$ $\neq0$ & $~~~$ $\neq0$ & $~~$
$\neq0$\\
\hline
\end{tabular}
\end{table}


\subsubsection{Correlated noise in the degradation rate}

We now consider the presence of correlated Gaussian noise which alters
the degradation rate in Eq.~(\ref{eq:stc}).  The corresponding
Langevin Eq.~(\ref{eq:stc}) can be rewritten in the form:
\begin{equation}\label{eq:mul}
\frac{dx}{dt} = r + a\frac{x^2}{1+x^2}-(1+\xi(t))x.
\end{equation}
Here the noise $\xi(t)$ is modulated due to the multiplication with
the state variable $x$.  Therefore, a small random fluctuation in the
degradation rate can lead to a sudden regime shift in the protein
concentration.  The role of noise intensity $\sigma_{1}$ and
correlation time $\tau_{1}$ of the noise $\xi(t)$ are very important
factors, because they can act as system parameters.  For fixed values
of the control parameters $r$ and $a$, changes in the noise intensity
$\sigma_{1}$ and the correlation time $\tau_{1}$ can trigger sudden
regime shifts in the level of protein concentration.

From Eq.~(\ref{Eq:Spdf2}), the SPDF corresponding to
Eq.~(\ref{eq:mul}) can be rewritten as:
\begin{eqnarray}
P_{s}(x) & = & \frac{N_{c}}{B(x)} \exp \Big[\int^{x}\frac{A(u)}{B(u)} du\Big]\nonumber\\
 & = & \frac{N_{c}}{\frac{\sigma_{1}x^2}{1-\tau_{1}f'(x_{s})}}
\times \exp\left[\int^x\frac{f(u)+\frac{\sigma_{1}u}{1-\tau_{1}f'(x_{s})}}{\frac{\sigma_{1}u^2}{1-\tau_{1}f'(x_{s})}}du\right].\nonumber\\
\label{Eq:Spdf1}
\end{eqnarray}
The potential function is derived from Eq.~(\ref{Eq:POT1}) and is
given by:
\begin{equation}\label{Eq:POT:DRate}
\phi(x)=\frac{1}{2} \ln \left[\frac{\sigma_{1}x^2}{1-\tau_{1}f^{'}(x_{s})}\right]-\int^x\frac{f(u)}{\frac{\sigma_{1}u^2}{1-\tau_{1}f'(x_{s})}}du.
\end{equation}

\begin{figure}[!h]
\begin{center}
\includegraphics[width=8.5cm]{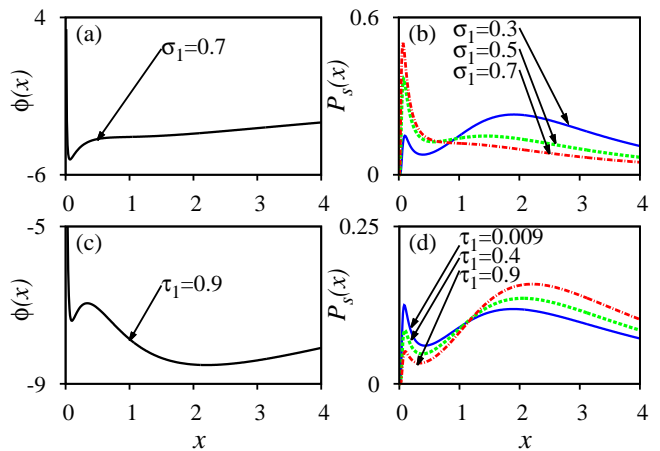}
\caption{\label{Fig:mul} (Color online) Stochastic potential $\phi(x)$
  and SPDF $P_{s}(x)$ of the system (\ref{eq:mul}): (a) $\phi(x)$ for
  the noise intensity $\sigma_{1}=0.7$, and (b) $P_{s}(x)$ for three
  different values of noise intensity $\sigma_{1}=0.3$ (blue curve),
  $\sigma_{1}=0.5$ (green curve) and $\sigma_{1}=0.7$ (red curve) with
  fixed $\tau_{1}=0.5$, $r=0.1$ and $a=3\:$. (c) $\phi(x)$ for the
  correlation time $\tau_{1}=0.9$, and (d) $P_{s}(x)$ for three
  different correlation time $\tau_{1}=0.009$ (blue curve),
  $\tau_{1}=0.4$ (green curve) and $\tau_{1}=0.9$ (red curve) with
  fixed $\sigma_{1}=0.5$, $r=0.1$ and $a=3.5\:$.  The increase in
  $\sigma_{1}$ induces regime shift from high to low protein
  concentration state, whereas increase in correlation time $\tau_{1}$
  induces regime shift from low to high protein concentration state.}
\end{center}
\end{figure}

\begin{figure}[!h]
\begin{center}
\begin{tabular}{l}
\resizebox{!}{2.0in}{\includegraphics{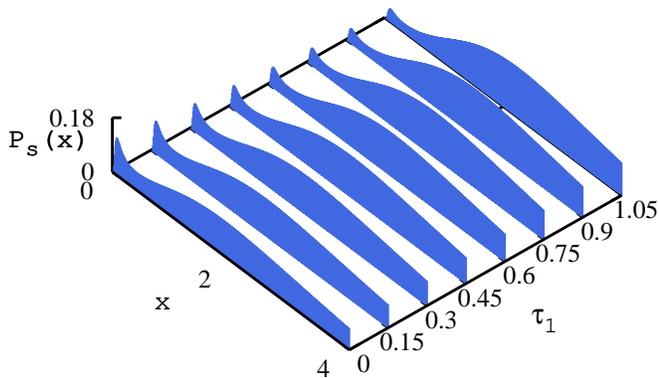}} \\
\end{tabular}
\caption{\label{steady1} (Color online) The evolution of SPDF $P_s(x)$
  of the system (\ref{eq:mul}) for continuously changing the
  correlation time $\tau_{1}$. The other parameters are
  $\sigma_{1}=0.5$, $r=0.1$ and $a=3.5$.  As the $P_s(x)$
  corresponding the right potential well has increased with increase
  in the $\tau_{1}$, the system experiences a regime shift from low to
  high protein concentration state.}
\end{center}
\end{figure}

The role of correlated noise on the relative stability between two
alternative steady states can be well understood by illustrating the
SPDF (\ref{Eq:Spdf1}) and the potential (\ref{Eq:POT:DRate}) for an
exemplary set of parameters.  Figures~\ref{Fig:mul}(a)-(b) show the
influence of the colored noise intensity $\sigma_{1}$ on the shape of
the potential $\phi(x)$ and the SPDF $P_{s}(x)$.  It can be seen that
for a fixed value of $\tau_{1}$, increasing values of $\sigma_{1}$
entail an increase in the likelihood of undesired regime shifts from
one stable state to another stable state (Fig.~\ref{Fig:mul}(a)).
With increasing values of $\sigma_1$, the SPDF peak at the low protein
concentration $x$ is increasing and that of the high protein
concentration $x$ is decreasing.  Hence, an increase in the noise
intensity $\sigma_{1}$ can induce a sudden regime shift from high to
low protein concentration state.  However, the dynamic correlation
time $\tau_{1}$ has inverted effect on the steady states of the system
(Figs.~\ref{Fig:mul}(c)--(d)).  Figure~\ref{Fig:mul}(d) depicts the
changes in the SPDF $P_{s}(x)$ peaks with changes in $\tau_{1}$ for a
fixed value of $\sigma_1$.  It is evident from the $P_{s}(x)$ peaks
that at low values of $\tau_{1}$ the lower state is more stable and at
high values of $\tau_{1}$ the upper state becomes more stable.  In
fact, $\tau_{1}$ has nontrivial effect on the stationary state and an
increase in $\tau_{1}$ can cause a regime shift form low to high
protein concentration state.  The above results indicate that
probability of shifting to the lower stable state is more in the case
of increasing noise intensity $\sigma_{1}$, whereas probability of
finding upper stable state is more in the case of increasing
correlation time $\tau_{1}$.  Figure~\ref{steady1} shows the
continuous evolution of the SPDF $P_s(x)$ with increasing values
$\tau_{1}$.

\begin{figure}[!h]
\begin{center}
\begin{tabular}{l}
\resizebox{!}{1.75in}{\includegraphics{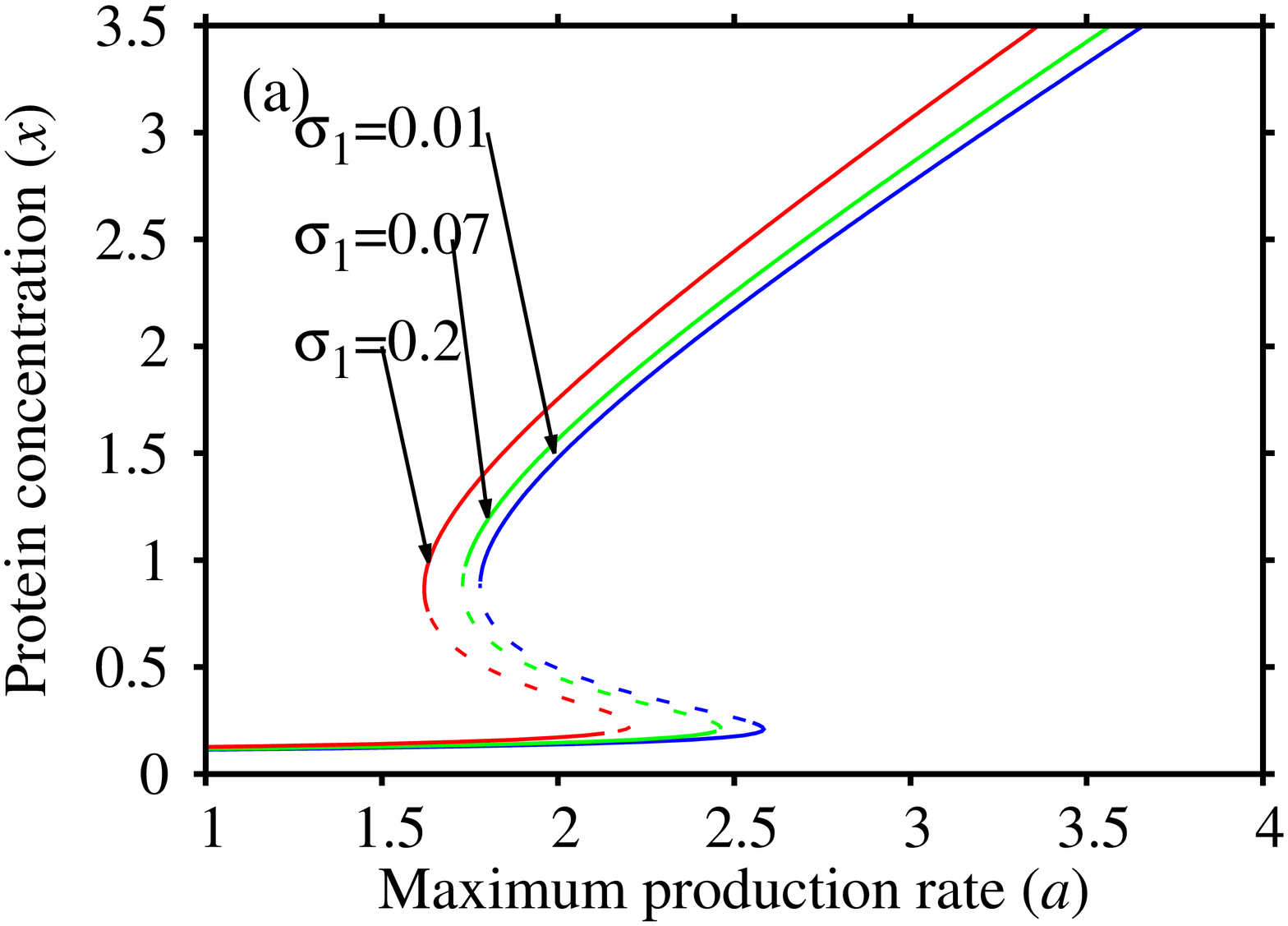}}\\
\resizebox{!}{1.75in}{\includegraphics{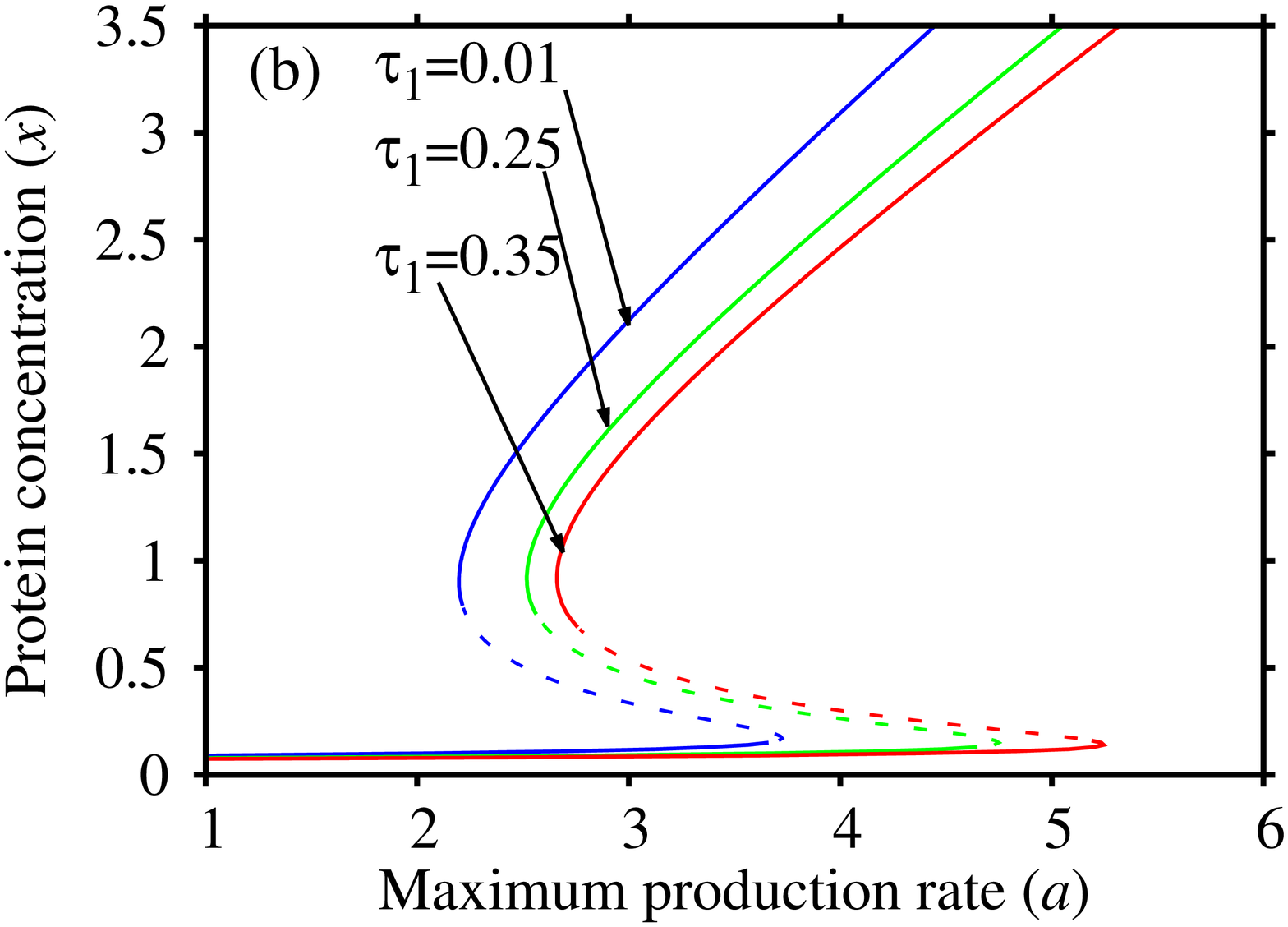}}\\
\end{tabular}
\caption{\label{Fig:bif1} (Color online) Extrema of the SPDF $P_s(x)$
  of the gene regulation model (\ref{eq:mul}) coupled with noise only
  in the degradation rate, as a function of $a$: (a) For a fixed
  $\tau_1=1.9$ and increasing values of $\sigma_{1}$, and (b) for a
  fixed $\sigma_{1}=0.2$ and increasing values $\tau_{1}$. The other
  parameters are $r=0.1$, $\sigma_{2}=0$, $\tau_{2}=0$, $\tau_{3}=0$
  and $\lambda=0$. The increase in $\sigma_{1}$ reduces the
  bistability regime, whereas increase in $\tau_{1}$ increases the
  bistabiltiy regime.}
\end{center}
\end{figure}

From Eq.~(\ref{Eq:Ext}), now the extrema of $P_s(x)$ can be written
as:
\begin{equation}\label{Eq:Ext:DRate}
f(x)-\frac{\sigma_{1}x}{1-\tau_{1}f'(x_{s})}=0.
\end{equation}
Using the above Eq.~(\ref{Eq:Ext:DRate}), the extrema of $P_s(x)$ is
plotted in Fig.~\ref{Fig:bif1} as a function of the maximum
transcription rate $a$.  With the help of the extrema, we investigate
the occurrence of critical transition (i.e., any qualitative changes
in the stationary state) in the stochastic system (\ref{eq:stc}) by
changing the noise intensity $\sigma_{1}$ and the correlation time
$\tau_{1}$. Changes in $\sigma_{1}$ and $\tau_{1}$ have opposite
effects on the steady state behavior of the system.  For a fixed
$\tau_{1}$, increasing values of $\sigma_{1}$ decreases the
bistability regime (Fig.~\ref{Fig:bif1}(a)) and for a fixed
$\sigma_{1}$, increasing values $\tau_{1}$ increases the bistability
regime (Fig.~\ref{Fig:bif1}(b)).

\subsubsection{Correlated noise in the basal rate}

We now focus only on the effect of correlated noise source $\eta(t)$
in the basal rate in Eq.~(\ref{eq:stc}) with stochastic parameters
$\sigma_{2}\neq 0$ and $\tau_{2}\neq 0$ (see Table~I).  In this case,
Eq.~(\ref{Eq:Spdf2}) can be written as:
\begin{eqnarray} \label{Eq:SPDF:NB}
P_{s}(x) & = & \frac{N_{c}}{B(x)} exp \Big[\int^{x}\frac{A(u)}{B(u)} du\Big]\nonumber\\
 & = & \frac{N_{c}}{\frac{\sigma_{2}}{1-\tau_{2}f'(x_{s})}}
\exp\left[\int^x\frac{f(u)}{\frac{\sigma_{2}}{1-\tau_{2}f'(x_{s})}}du\right],\nonumber\\
\end{eqnarray}
and the potential function is derived from Eq.~(\ref{Eq:POT1}) is
given by:
\begin{equation} \label{Eq:POT:NB}
\phi(x)=\frac{1}{2}\ln\left[\frac{\sigma_{2}}{1-\tau_{2}f'(x_{s})}\right]
-\int^x\frac{f(u)du}{\frac{\sigma_{2}}{1-\tau_{2}f'(x_{s})}}.
\end{equation}

\begin{figure}[!h]
\begin{center}
\includegraphics[width=8.5cm]{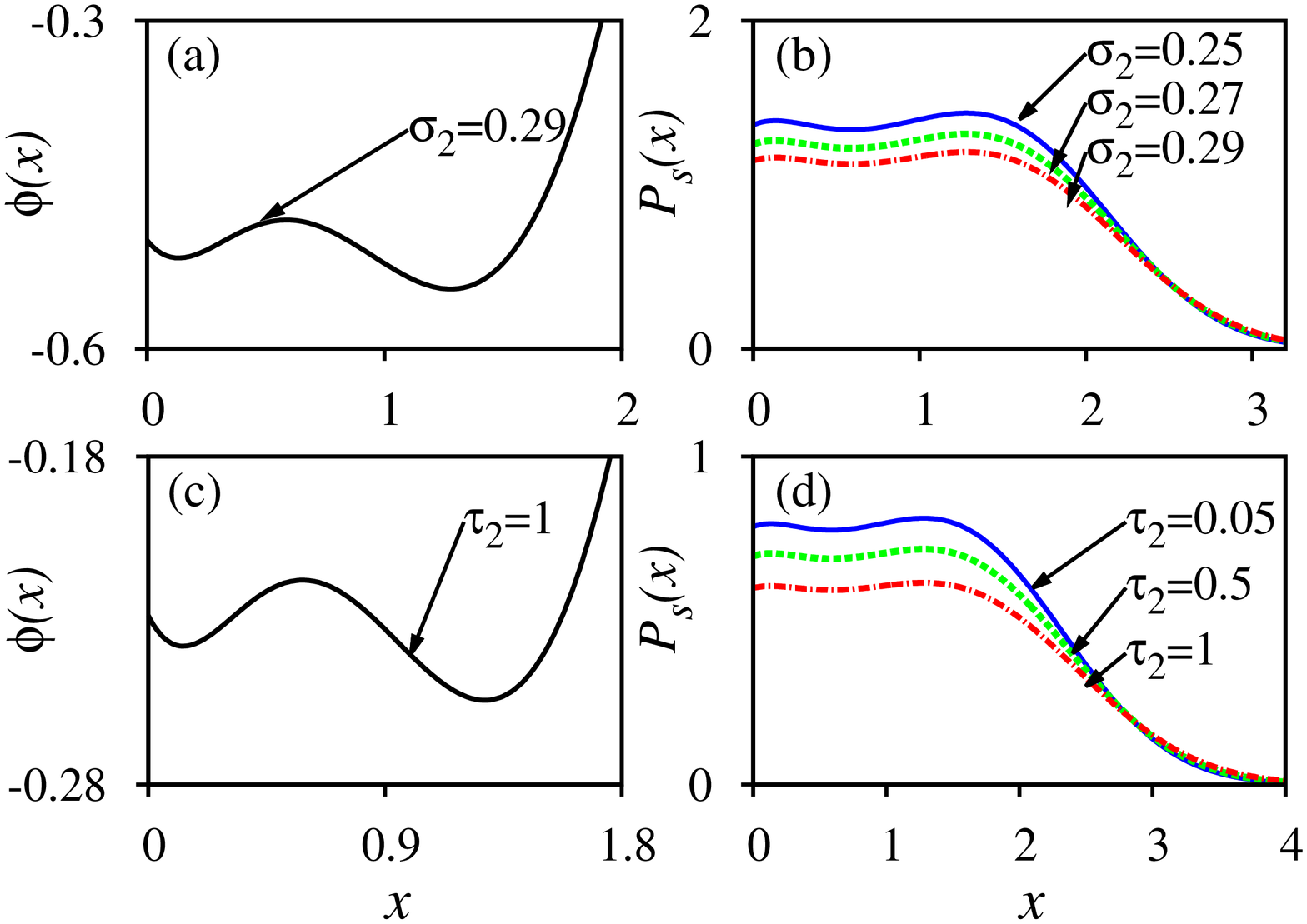}
\caption{\label{Fig:add} (Color online) Stochastic potential $\phi(x)$
  and SPDF $P_{s}(x)$ when noise is present only in the basal rate:
  (a) $\phi(x)$ for the noise intensity $\sigma_{2}=0.29$, and (b)
  $P_{s}(x)$ for three different values of the noise intensity
  $\sigma_{2}=0.25$ (blue curve), $\sigma_{2}=0.27$ (green curve) and
  $\sigma_{2}=0.29$ (red curve) with fixed $\tau_{2}=1$, $r=0.1$ and
  $a=1.9$.  (c) $\phi(x)$ for the correlation time $\tau_{2}=1$, and
  (d) $P_{s}(x)$ for three different of the correlation time
  $\tau_{2}=0.05$ (blue curve), $\tau_{2}=0.5$ (green curve) and
  $\tau_{2}=1$ (red curve) with fixed $\sigma_{2}=0.5$, $r=0.1$ and
  $a=1.9$.  The increase in $\sigma_{2}$ and $\tau_{2}$ has not much
  effect on $\phi(x)$ and $P_{s}(x)$, however, both the valleys and
  tops of $P_{s}(x)$ decay in height.}
\end{center}
\end{figure}

\begin{figure}[!h]
\begin{center}
\includegraphics[width=8.5cm]{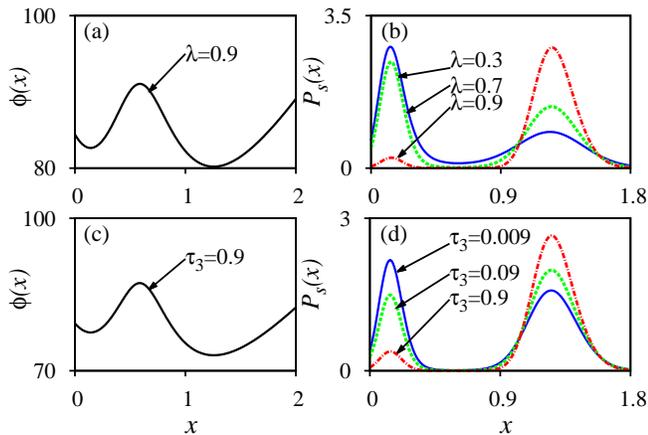}
\caption{\label{Fig:mul1} (Color online) Stochastic potential
  $\phi(x)$ and SPDF $P_{s}(x)$, when the model (\ref{eq:stc}) driven
  by cross-correlated noises: (a) $\phi(x)$ for the cross-correlation
  strength $\lambda=0.9$, and (b) $P_{s}(x)$ for three different
  values of the cross-correlation strength $\lambda=0.3$ (blue curve),
  $\lambda=0.7$ (green curve) and $\lambda=0.9$ (red curve) with
  $a=1.9$, $r=0.1$, $\sigma_{1}=0.01$, $\sigma_{2}=0.004$,
  $\tau_{1}=0.01$, $\tau_{2}=0.08$ and $\tau_{3}=0.03$.  (c) $\phi(x)$
  for the correlation time $\tau_{3}=0.9$, and (d) $P_{s}(x)$ for
  three different values of the correlation time $\tau_{3}=0.009$
  (blue curve), $\tau_{3}=0.09$ (green curve) and $\tau_{3}=0.9$ (red
  curve) with $a=1.9$, $r=0.1$, $\sigma_{1}=0.01$, $\sigma_{2}=0.004$,
  $\lambda=0.3$, $\tau_{1}=0.01$ and $\tau_{2}=0.08$.  The increase in
  both the $\lambda$ and $\tau_{3}$ induce regime shifts from low to
  high protein concentration state.}
\end{center}
\end{figure}

Figure~\ref{Fig:add} depicts the stochastic potential $\phi(x)$ and
SPDF $P_{s}(x)$ for different values of the noise intensity
$\sigma_{2}$, and the noise correlation time $\tau_{2}$.  We set the
parameters in such a way that the system is in the bistable regime,
i.e., both the high and low protein concentration states.  Our results
show that for a fixed $\tau_{2}$ increasing values of $\sigma_{2}$
have equal effect on the relative stability of both the steady states
(Figs.~\ref{Fig:add}(a)--(b)).  The same result follows for fixed
$\sigma_{2}$ and increasing values of $\tau_{2}$
(Figs.~\ref{Fig:add}(c)--(d)).  What we find is that the bimodal
distribution of $\phi(x)$ and $P_{s}(x)$ are retained, and the
positions of the steady states also remains almost the same, however
the valleys and the tops of $P_{s}(x)$ decay in height with increase
in both $\sigma_{2}$ and $\tau_{2}$.


\subsubsection{Correlated noise in both the basal and 
degradation rate with cross-correlation strength $\lambda$}

In this section, we consider the Langevin Eq.~(\ref{eq:stc}) in the
presence of both the colored noises $\xi (t)$ and $\eta (t)$.
Furthermore, $\xi (t)$ and $\eta (t)$ are statistically cross
correlated with the cross-correlation strength $\lambda$.  The cross
correlation between $\xi (t)$ and $\eta (t)$ is chosen due to the
regulation of feedback mechanism, i.e., in the presence of noise the
protein concentration $x$ is chemically coupled to the degradation
rate \cite{Ro2010}.  Here, our goal is to understand the impact of the
cross-correlation strength $\lambda$ and correlation time $\tau_{3}$
between two noises $\xi (t)$ and $\eta (t)$, on the steady states of
the system and the transition between them.

\begin{figure}[!h]
\hspace{-8cm}(a) 
\begin{center}
\begin{tabular}{l}
(b)
\resizebox{!}{1.85in}{\includegraphics{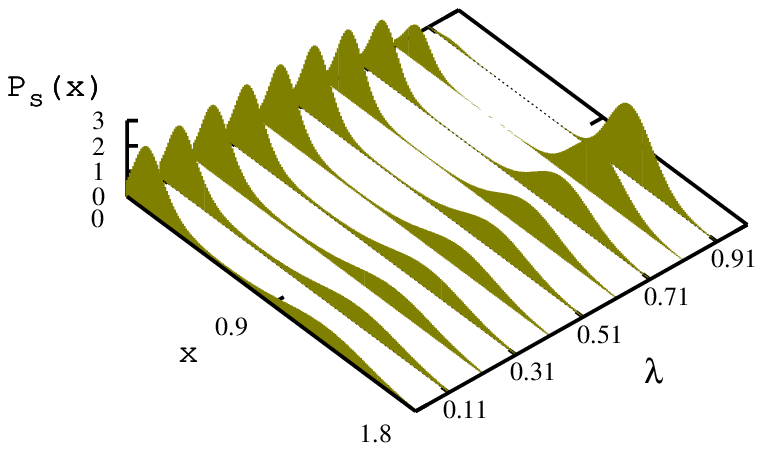}} \\
\resizebox{!}{1.9in}{\includegraphics{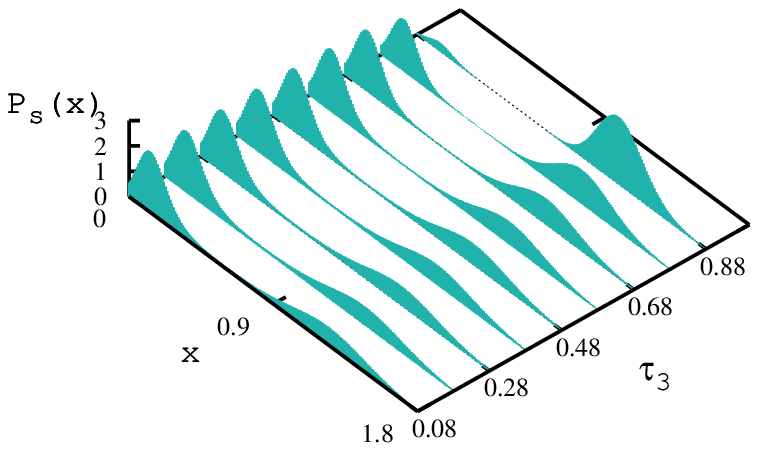}} \\
\end{tabular}
\caption{\label{3D_CCN} (Color online) The evolution of SPDF $P_s(x)$
  of the system (\ref{eq:mul}) with continuous changes: (a) in the
  cross-correlation strength $\lambda$ for $\tau_{3}=0.03$, and (b) in
  the correlation time $\tau_{3}$ for $\lambda=0.3$.  The other
  parameters are $r=0.1$, $a=1.9$, $\sigma_{1}=0.01$,
  $\sigma_{2}=0.004$, $\tau_{1}=0.01$ and $\tau_{2}=0.08$.}
\end{center}
\end{figure}

Using Eqs.~(\ref{Eq:Spdf2}) and (\ref{Eq:POT1}) we compute the SPDF
$P_{s}(x)$ and the potential function $\phi(x)$ for the system
(\ref{eq:stc}).  Figures~\ref{Fig:mul1}(a)-(b) show the radical effect
of the cross-correlation strength $\lambda$ on the shape of $\phi(x)$
and $P_{s}(x)$.  For a fixed value of $\tau_{3}$, with increasing
values of $\lambda$, the SPDF peak at low protein concentration state
is reducing and that of high protein concentration state is increasing
(see Fig.~\ref{Fig:mul1}(b) for $\lambda=0.9$).  Hence, an increase in
$\lambda$ can induce a sudden regime shift from low protein
concentration state to high protein concentration state.

Moreover, the correlation time $\tau_{3}$ has similar effect on the
shape of $\phi(x)$ and $P_{s}(x)$ likewise the effect of
cross-correlation strength $\lambda$ (Figs.~\ref{Fig:mul1}(c)-(d)).
It is evident from the $P_{s}(x)$ peak that at low value of
$\tau_{3}$, the lower steady state is more stable in comparison with
the higher steady state, whereas at high value of $\tau_{3}$, the
scenario is just opposite (Fig.~\ref{Fig:mul1}(d)).  The above results
indicate that probability of shifting to the upper steady state is more
for both the cases: Increasing the cross-correlation strength
$\lambda$ and the correlation time $\tau_{3}$ \cite{Frcasaib12}.
Figures~\ref{3D_CCN}(a)-(b) show the continuous evolution of the SPDF
$P_{s}(x)$ with increasing values of $\lambda$ and $\tau_{3}$.
  
\begin{figure}[!h]
\begin{center}
\begin{tabular}{l}
\resizebox{!}{1.75in}{\includegraphics{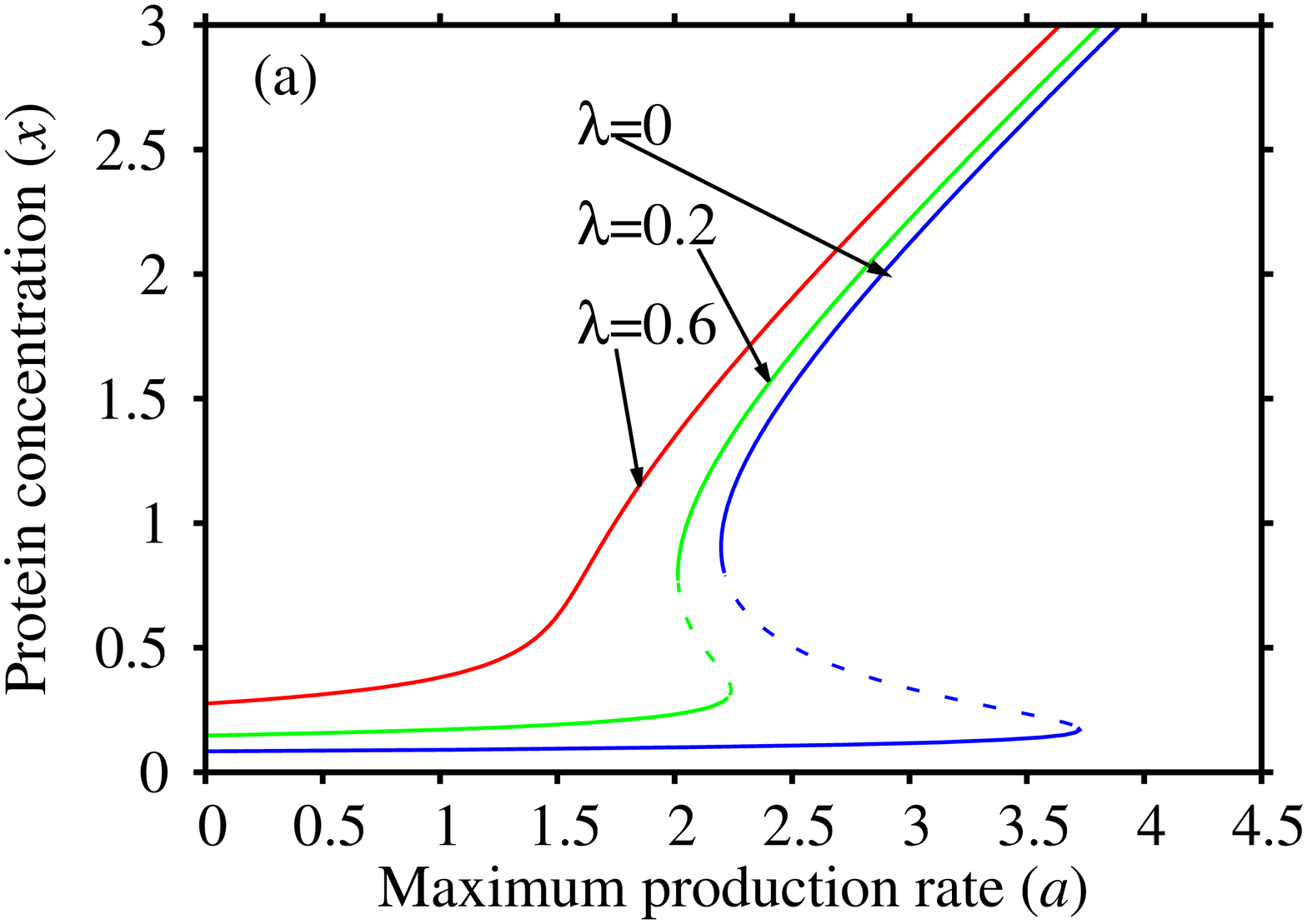}}\\
\resizebox{!}{1.75in}{\includegraphics{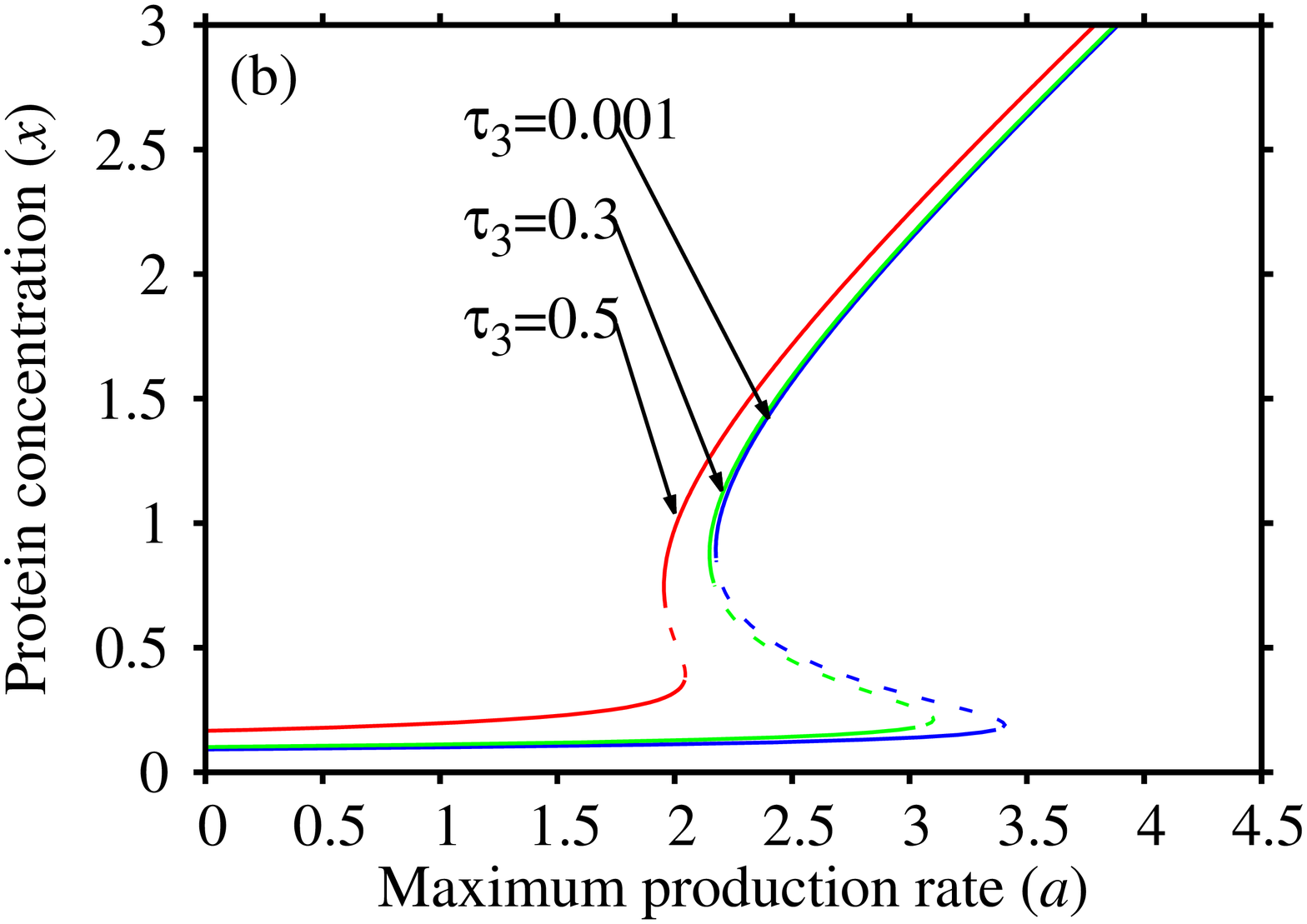}}\\
\end{tabular}
\caption{\label{Fig:bif_PD} (Color online) Extrema of the SPDF
  $P_s(x)$ of the gene regulation model (\ref{eq:stc}) driven by
  cross-correlated noises, as a function of $a$: (a) For increasing
  values of the cross-correlation strength $\lambda$ with other
  parameter values are $r=0.1$, $\sigma_{1}=0.2$, $\sigma_2=0.5$,
  $\tau_{1}=0.01$, $\tau_{2}=0.01$ and $\tau_{3}=0.1$, and (b) for
  increasing values of the correlation time $\tau_{3}$ with other
  parameter values are $r=0.1$, $\sigma_{1}=0.2$, $\sigma_{2}=0.5$,
  $\tau_{1}=0.5$, $\tau_{2}=0.5$ and $\lambda=0.1$.  The bistability
  regime reduces with increase in both $\lambda$ and $\tau_{3}$.}
\end{center}
\end{figure}
  
Using Eq.~(\ref{Eq:Ext}), the extrema of SPDF $P_{s}(x)$ is depicted
in Figs.~\ref{Fig:bif_PD}(a)-(b) as a function of the maximum
transcription rate $a$.  Notice that, with increasing values of
$\lambda$ and $\tau_{3}$ both extrema curves exhibit similar behavior.
As an example, Fig.~\ref{Fig:bif_PD}(a) shows that increase in
$\lambda$ between two noises reduce the bistability region and for
higher values of $\lambda$, bistability completely disappears.  These
results indicate that correlated stochastic fluctuations in gene
regulation can significantly effect the bistable states and even it can
reduce it to monostable state.  Moreover, the relative stability of the
bistable states are dynamically coupled with the correlation
parameters of the noise.

\subsection{Mean first-passage time of the system driven
 by cross-correlated noises}\label{sec:3c}

For stochastic bistable systems, it is important to estimate the
amount of time between shifts from one steady state to another steady
state.  As it helps to quantify the effects of noise on the regime
switching between alternative steady states.  This time is often
referred as first-passage time.  When the first-passage time is
averaged over many realizations, the resulting time is called mean
first-passage time (MFPT) \cite{Ga85}.  To examine the robustness of
steady states, MFPT provides a very useful characterization.  A longer
MFPT implies the state is more stable.  Now, we study the influence of
cross-correlation strength $\lambda$ and correlation time $\tau_{3}$
on the MFPT.

\begin{figure}[!h]
\begin{center}
\begin{tabular}{l}
\resizebox{!}{3.8in}{\includegraphics{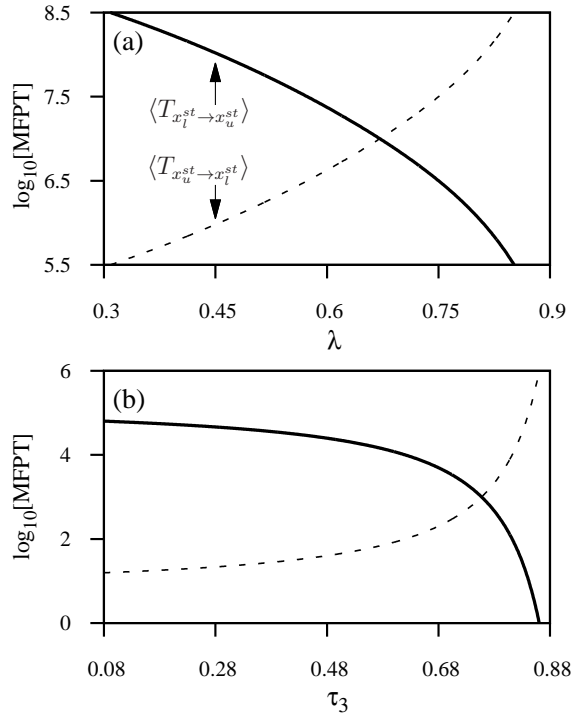}}\\
\end{tabular}
\caption{\label{Fig:bbb} (Color online) The effect of $\lambda$ and
  $\tau_{3}$ on the MFPT.  (a) The MFPT $\langle T_{x_{l}^{st}
    \rightarrow x_{u}^{st}} \rangle$ decreases and $\langle
  T_{x_{u}^{st} \rightarrow x_{l}^{st}} \rangle$ increases with the
  increase of $\lambda$ for $\tau_{3}=0.03$.  (b) Similar situation
  arises with the increase of $\tau_{3}$ for $\lambda=0.3$.  The other
  parameters are same as in Fig.~\ref{3D_CCN}. }
\end{center}
\end{figure}

To start with, let $x_{l}^{st}$ be the low and $x_{u}^{st}$ be the
high protein concentration states, separated by a potential barrier
$x_{b}^{un}$ (working as a basin boundary between the two steady
states $x_{l}^{st}$ and $x_{u}^{st}$) of the system (\ref{eq:stc}).
The basin of attraction of the state $x_{u}^{st}$ extends from
$x_{b}^{un}$ to $+\infty$, as it is in the right of $x_{l}^{st}$.  The
MFPT $\langle T(x) \rangle$, can be obtained by solving the following
ordinary differential equation \cite{Ga85}:
\begin{equation}\label{mean}
A(x)\frac{\partial \langle  T \rangle}{\partial x}+\frac{1}{2}B(x)\frac{\partial^{2} \langle T \rangle}{\partial x^{2}}=-1,
\end{equation}
with boundary conditions $\langle T(x^{un}_b) \rangle = 0$ and
$\frac{\partial \langle T (+\infty) \rangle }{\partial x}=0$, where
$A(x)$ and $B(x)$ are respectively given by Eqs.~(\ref{eq:exp1}a)
and~(\ref{eq:exp1}b).


\begin{figure*}[!ht]
\begin{center}
\includegraphics[width=1.5\columnwidth]{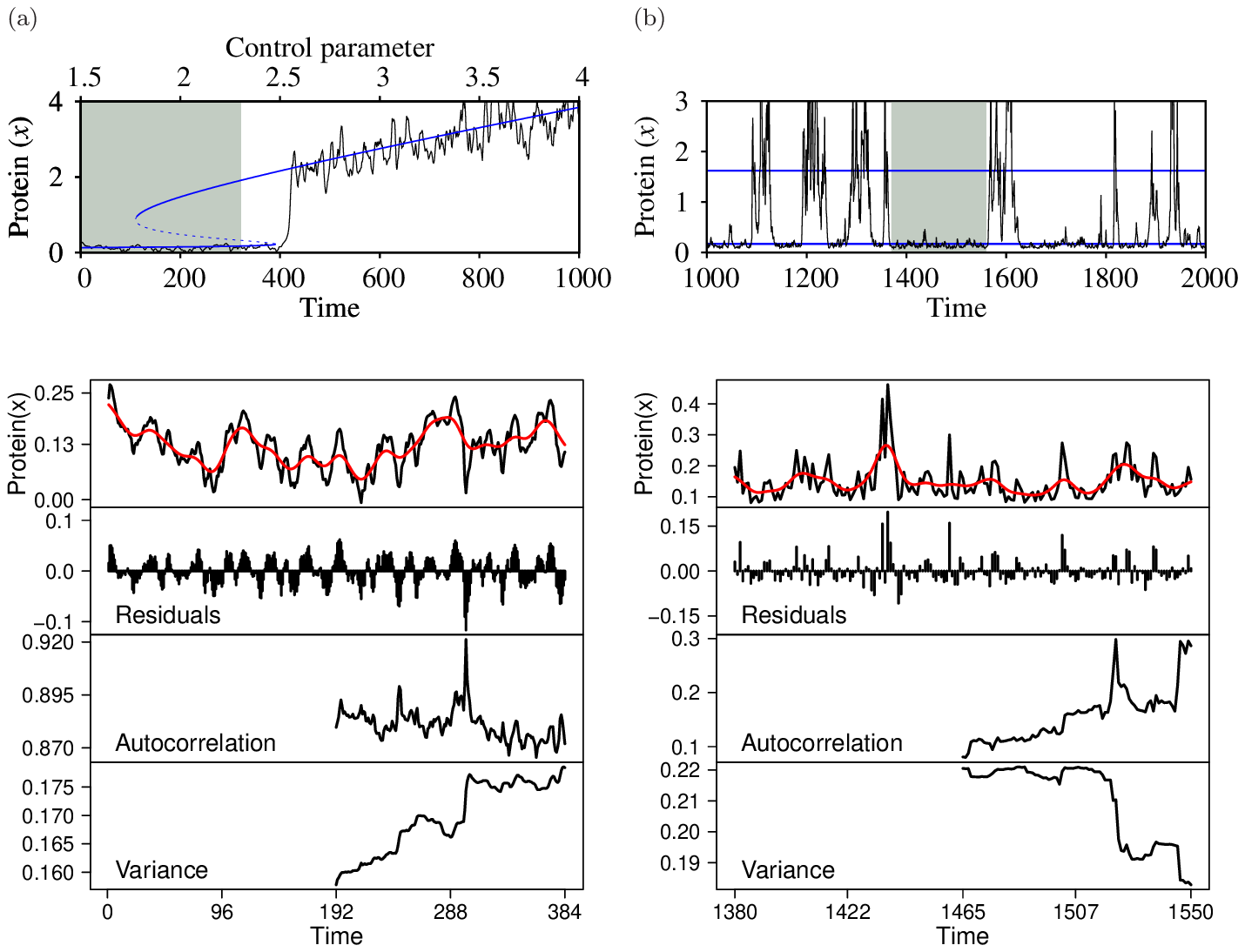}
\caption{\label{csd} (Color online) Early warning signals for
  simulated time series data of the stochastic model in the case of:
  (a) CSD and (b) SS.  The variance and autocorrelation are calculated
  using moving window of half the length of the time series segments
  (segments are indicated by the shaded regions): (a) For CSD:
  parameter values are $r=0.1$, $\sigma_{1}=0.002$, $\sigma_{2}=0.09$,
  $\lambda=0.8$, $\tau_{1}=5$, $\tau_{2}=5$ and $\tau_{3}=1$; (b) For
  SS: parameter values are $r=0.1$, $a=1.9$, $\sigma_{1}=0.005$ and
  $\sigma_{2}=0.007$, $\lambda=0.01$, $\tau_{1}=0.09$, $\tau_{2}=0.09$
  and $\tau_{3}=1$.  The increase in variance act as a robust
  indicator for CSD, whereas variance fails in the case of SS. The
  autocorrelation gives weak trend in both CSD and SS.}
\end{center}
\end{figure*}


By solving the Eq.~(\ref{mean}), we obtain the expressions of MFPT for
$x_{l}^{st}$ and $x_{u}^{st}$.  The expressions for MFPT $\langle
T_{x_{l}^{st} \rightarrow x_{u}^{st}} \rangle$ and $\langle
T_{x_{u}^{st} \rightarrow x_{l}^{st}} \rangle$ are given by
\cite{Ga85}:
\begin{eqnarray}
\langle T_{x_{l}^{st} \rightarrow
    x_{u}^{st}} \rangle & = & 2 \int_{x_{l}^{st}}^{x_{b}^{un}}\frac{dy}{\omega(y)}\int_{0}^{y}\frac{\omega(z)}{B(z)}dz, \;\; \text{and}\\
\langle T_{x_{u}^{st} \rightarrow
    x_{l}^{st}} \rangle  & = & 2 \int_{x_{b}^{un}}^{x_{u}^{st}}\frac{dy}{\omega(y)}\int_{y}^{\infty}\frac{\omega(z)}{B(z)}dz,
\end{eqnarray}
where
\begin{equation*}
w(x)= \exp \Big(\int_{x_{0}}^{x}\frac{2A(u)}{B(u)}du\Big),
\end{equation*}
with $x_{0}=0$ for the $x_{l}^{st}\rightarrow x_{u}^{st}$ transition
and $x_{0}=x_{b}^{u_{n}}$ for the $x_{u}^{st}\rightarrow x_{l}^{st}$
transition.

Effects of changing $\lambda$ and $\tau_{3}$ on the MFPT are shown in
Fig.~\ref{Fig:bbb}.  We found that the MFPT $\langle T_{x_{l}^{st}
  \rightarrow x_{u}^{st}} \rangle$ decreases and $\langle
T_{x_{u}^{st} \rightarrow x_{l}^{st}} \rangle$ increases, with
increase in the cross-correlation strength $\lambda$
(Fig.~\ref{Fig:bbb}(a)).  Hence, an increase in $\lambda$ results in a
regime shift from the left potential well (low concentration state of
$x$) to the right potential well (high concentration state of $x$).
We observe similar dynamics with variations in $\tau_3$
(Fig.~\ref{Fig:bbb}(b)).  The conclusions drawn from the analysis of
MFPT are also consistent with the SPDF $P_s(x)$ shown in
Fig.~\ref{3D_CCN}.  This result highlights the significance of
correlated noise in gene expression dynamics.


\subsection{Precursors of regime shift}\label{sec:3d}

Here, the main emphasis is to explore the robustness of EWS (e.g.,
lag-$1$ autocorrelation, variance and conditional heteroskedasticity)
as indicators of regime shifts in protein concentration levels.  In
clinical medicine, EWS can be considered as bio-markers because these
are indicators of regime shifts in biological state for living
organism \cite{ChLi12}.  However, earlier techniques or bio-markers are
mainly used to investigate the current disease state of an organ based
on metabolites or individual protein level \cite{Cui2011,Jin2009}.

For our analysis, we consider stochastic time series of the model
(\ref{eq:stc}) for both the cases, critical slowing down (CSD) and
stochastic switching (SS)
\cite{ScJo09,Scheffer:2012sc,Dakos:2012pone,sc2015}.  The presence of
cross-correlated noise in the degradation and basal rates are
considered.  Numerical simulations have been performed using the
Euler-Maruyama method \cite{DH01} with an integration step-size of
$0.001$.  In the time series, we first visually identify shifts
between low to high protein concentration.  Then we took time series
segments (the shaded regions in Fig.~\ref{csd}) prior to a regime
shift and analyze them for the presence of EWS.  For stationarity in
residuals, we used Gaussian detrending with bandwidth 40, before
performing any statistical analysis of the data.  Then we used a
moving window size of half the length of the considered time series
segment.  The time series analysis have been performed using the
``Early Warning Signals Toolbox''
(http://www.early-warning-signals.org/).  First, we calculate the
variance and lag-$1$ autocorrelation, as these two indicators are
known to be most appropriate to anticipate regime shifts.  The
autocorrelation at lag-1 is given by the autocorrelation function
(ACF): $\displaystyle
\rho_{1}=\frac{E\left[(x(t)-\mu)(x(t+1)-\mu)\right]}{\sigma^{2}}$,
where $E$ is the expected value operator, $x(t)$ is the value of the
state variable at time $t$, and $\mu$ and $\sigma^{2}$ are the mean
and variance of $x(t)$, respectively.  Variance is the second moment
around the mean $\mu$ and measured as: $\displaystyle
\sigma^{2}=\frac{1}{N}\Sigma_{i=1}^{N}(x(t)-\mu)^{2}$, where $N$ is
the number of observations within the considered moving window.  A
concurrent rise in these indicators forewarn an upcoming regime shift
\cite{Book_Sch,ScJo09}.


\begin{figure}
\begin{center}
\includegraphics[width=0.99\columnwidth]{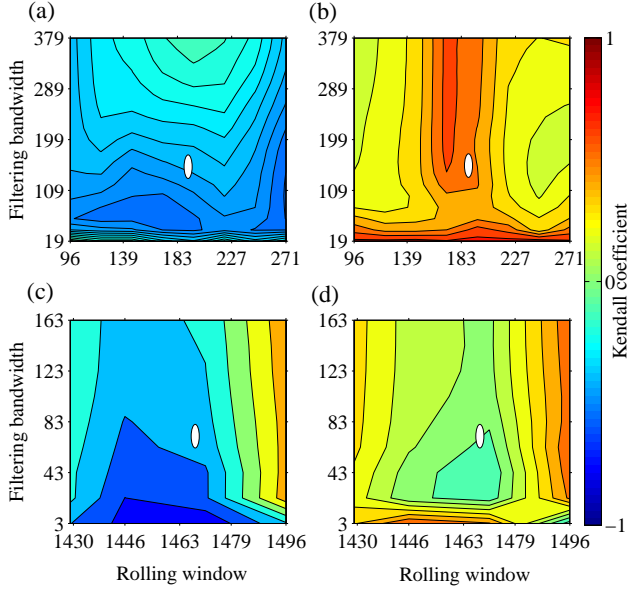}
\caption{\label{Fig:sens}(a) (Color online) Contour plots depict the
  effects of moving window size and filtering bandwidth on the
  measured autocorrelation ((a) and (c)) and variance ((b) and (d))
  for the: (a-b) CSD data and (c-d) SS data shown in Fig.~\ref{csd} as
  estimated by the Kendall's coefficient.  The empty ovals indicate
  the choices of the window size and filtering bandwidth used in the
  calculations in Fig.~\ref{csd}.}
\end{center}
\end{figure}


\begin{figure}[!h]
\begin{center}
\includegraphics[width=0.85\columnwidth]{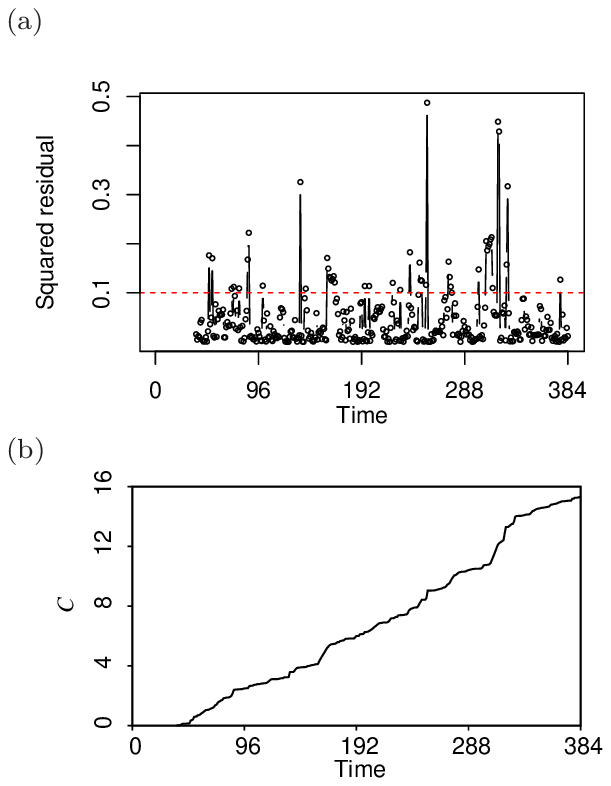}
\caption{\label{Fig:b2} (Color online) CH and cumulative numbers of
  test estimated on the CSD dataset prior to a regime shift.  (a) The
  squared residuals above the dashed red line are significant.  Here,
  the red line represents the level of significance.  (b) The
  cumulative number $(C)$ of significant Lagrange multiplier test
  applied to the time series obtained from the model.  The number
  $(C)$ increases prior to the transition indicating that significant
  number of tests shows conditional heteroskedasticity.}
\end{center}
\end{figure}


Figure~\ref{csd}(a) shows increase in $\sigma^2$ and decrease in
$\rho_1$ before a regime shift for the case of CSD.  Hence, in this
case $\sigma^2$ is able to successfully detect a regime shift in
protein concentration, whereas $\rho_1$ fails. However, in the case of
SS (Fig.~\ref{csd}(b)), both of these indicators fails.  For SS, the
failure of $\sigma^2$ and $\rho_1$ as EWS is in agreement with the
previous studies \cite{Drake:2013,Boet:2013,SKP14,Sh2016}.  The result
of EWS analysis also depends on the choice of factors like filtering
bandwidth and moving window size, used to calculate the standard
deviation and autocorrelation \cite{Dakos:2012pone}.  Hence, it is
important to investigate the robustness of our results with respect to
the choice of these factors.  In particular, we perform sensitivity
analysis which is necessary for the selection of bandwidth and moving
window size to maximize the estimated trend of EWS.  For CSD, we
estimate variance and autocorrelation in window size ranging from
$25\%$ to $71\%$ (i.e., $96$ to $271$ data points) of the time series
length, and for filtering bandwidth ranging from $5\%$ to $100\%$ (see
Figs.~\ref{Fig:sens}(a) and \ref{Fig:sens}(b)).  For SS, we use window
size ranging from $25\%$ to $68\%$ (i.e., $1430$ to $1496$ data
points) and bandwidth ranging from $2\%$ to $100\%$ (see
Figs.~\ref{Fig:sens}(c) and \ref{Fig:sens}(d)).
Figures~\ref{Fig:sens}(a) and \ref{Fig:sens}(c) represent contour
plots of rolling window size verses bandwidth for the autocorrelation,
similarly Figs.~\ref{Fig:sens}(b) and \ref{Fig:sens}(d) for the
variance.  The empty ovals in Fig.~\ref{Fig:sens} indicate the values
those we have used to calculate EWS in Fig.~\ref{csd}.  It is clear
that the autocorrelation in both the cases CSD
(Fig.~\ref{Fig:sens}(a)) and SS (Fig.~\ref{Fig:sens}(c)) do not give
proper result due to the low value of Kendall's coefficient
\cite{Dakos:2012pone}.  However, increasing trend in variance is found
in the case of CSD due to the proper selection of window size and
bandwidth corresponding to the high value of Kendall's coefficient,
which is also evident from the Fig.~\ref{Fig:sens}(b).


\begin{figure}[!h]
\begin{center}
\includegraphics[width=0.85\columnwidth]{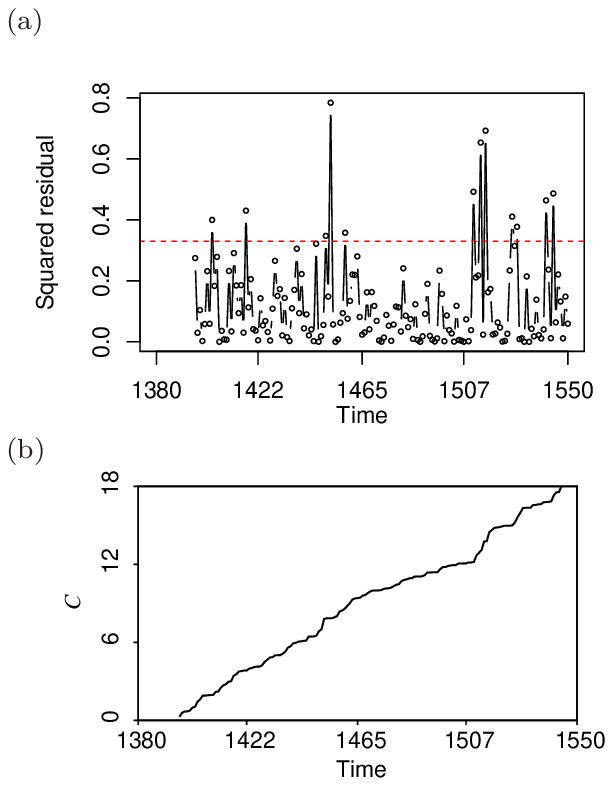}
\caption{\label{Fig:b21} (Color online) CH and cumulative numbers of
  test estimated on the SS dataset prior to regime shift.  (a) The
  points of CH above the dashed red line are significant.  Here, the
  red line represents the level of significance.  (b) The cumulative
  number $(C)$ of significant Lagrange multiplier test applied to the
  time series obtained from the model. The number $(C)$ increases
  prior to the transition indicating that significant number of tests
  shows CH.}
\end{center}
\end{figure}

Although autocorrelation and variance are known to be the most
preferred indicators to predict regime shifts, the fact is that they
are not always successful as shown in the previous examples.  This
arises because not all the regime shifts are associated with CSD
\cite{DaCa15}.  Moreover, improper data length, statistical
limitations and other types of transitions, such as purely
noise-induced transitions increase the risk of of false predictions.
We cannot avoid the possibility of false alarms completely
\cite{sc2015,DaCa15}.  Hence, we further tested another indicator
conditional heteroskedasticity (CH) (see Figs.~\ref{Fig:b2} and
\ref{Fig:b21}) \cite{Se11}.  CH is denoted by the persistence in the
conditional variance of the error terms.  In time series, it looks
like as cluster of high variability near a critical transition and
cluster of low variability far from the transition. This type of
clustering is known to be a leading indicator of regime shifts.  CH
provides threshold value for detecting regime shift and gives an
indication of upcoming regime shift \cite{Se11}.

We compute CH using moving window Lagrange multiplier test (window
width $10\%$ of the data) \cite{Se11}. First we extract the residuals
of a fitted model to the time series, then we fit an auto-regressive
model of selected order: 
\begin{equation*}
x_{t}=a_{0}+\sum_{i=1}^{q} a_{i}x_{t-i} + \epsilon_{t},
\end{equation*}
where the order $q$ is selected according to the Akaike information
criterion \cite{Ak1998} which is a measure of the relative goodness of
the fit.  Then we squared the residuals $\epsilon_{t}$, and finally
the residuals are regressed on themselves lagged by one time step:
\begin{equation*}
\epsilon^2_{t}=\alpha_{0}+\sum_{i=1}^{q} \alpha_{i} \epsilon^2_{t-i},
\end{equation*}
where $\alpha_{0}$ and $\alpha_{i}$ denotes the regression
coefficients.  The relationship between squared residuals
$\epsilon_t^2$ at lag-1 gives the properties of CH.  We also perform
chi square test to compare the values of squared residuals to a
$\chi^{2}$ distribution to identify the number of significant tests
where the CH is observed.  The cumulative number of significant tests
$(C)$ for CH applied to time series, is expected to increase as the
regime shift is approached.  Here Fig.~\ref{Fig:b2}(a)
(Fig.~\ref{Fig:b21}(a)) represents the CH estimated on the CSD (SS)
dataset prior to a regime shift which shows the positive relationship
of error variance and represents the significant CH (i.e., squared
residuals) above the significance level.  In Figs.~\ref{Fig:b2}(a) and
Fig.~\ref{Fig:b21}(a), the significance level is represented by the
red line. The residuals above this red line indicates that there is
presence of CH.  Figure~\ref{Fig:b2}(b) (Fig.~\ref{Fig:b21}(b)) shows
the result of cumulative number of significant tests $(C)$ for CH
applied to time series for the case of CSD (SS) and which is
increasing prior to a regime shift and gives positive EWS.  It is
important to observe that in the case of SS the indicator CH is
successful in comparison with autocorrelation and variance and this
is evident from Figs.~\ref{Fig:sens} and \ref{Fig:b21}.


\section{Discussion}\label{sec:dis}

Noise correlation can play a pivotal role in controlling the
regulatory functions of gene expression \cite{du2008,Si2006,Ro2005}.
In this paper, we have presented theoretical analysis and numerical
simulation of a gene expression model to study the role of Gaussian
colored noise in inducing sudden regime shifts at the levels of
protein concentration.  We have used the Ornstein-Uhlenbeck process
with the Langevin and Fokker-Plank descriptions to study the effects
of Gaussian colored noise.  Though one of our main goals is to
investigate the effects of noise correlation, for the sake of
completeness we also simultaneously studied the effects of colored
noise intensity.  The theoretical tools used to serve our purpose are
the stochastic potential, the stationary probability density function
and the mean first-passage time \cite{Ga85}.  For the presence of
colored noise in the protein degradation rate, we have shown that for
a fixed correlation time increase in the noise intensity induces
regime shift from high (``on'' state) to low (``off'' state) protein
concentration state.  Surprisingly, for a fixed noise intensity an
increase in the correlation time produces opposite result, it induces
regime shift from low (``off'' state) to high (``on'' state) protein
concentration state.  Moreover, with the help of the extrema of SPDF
we show that for a fixed correlation time, increasing values of noise
intensity reduces the bistability regime and for a fixed noise
intensity, increasing values of noise correlation increases the
bistability regime.  Our results also show that colored noise in the
basal rate retain the bimodal distribution of the steady states.  In
the case of cross-correlated colored noises in basal and degradation
rates, we have shown that both the cross correlation strength and
cross correlation time can induce regime shifts from low to high
protein concentration state, but reduce the bistable regime.  The
results of MFPT for cross-correlated colored noises also matches with
the outcome of stochastic potential and SPDF.  Thus, unlike earlier
studies on gene expression noise
\cite{Ha00,KaEl05,Frcasaib12,Gh12,Sh2016}, our findings suggest that
Gaussian colored noise can also induce sudden phenotypic variability
(i.e., regime shifts between ``on'' and ``off'' expression state) in
cells and the noise correlation time can act as a control parameter
for that.

Anticipation of regime shifts in gene expression could improve early
therapeutic intervention in complex human diseases
\cite{kor2014,gl15,Ri2016}.  Furthermore, EWS for predicting state
shifts in complex biological systems can be very useful as a bio-marker
for incurable and chronic human diseases where the stage of the
disease is an important factor of therapy and prognosis; for example
in liver cancer and lymphoma \cite{ChLi12}.  Keeping in mind the
complexity of cancer, if the stage of cancer can be identified by
using EWS, it would be remarkable.  Nonetheless, the success of EWS
in anticipating catastrophic shifts in ecosystem experiments
\cite{Wang:2012} suggests that it could be possible to develop and
employ EWS in cancer biology based on clinical trials
\cite{kor2014,TrAn15}.  Considering, both CSD and SS time series data
of the gene expression model we show that variance and autocorrelation
sometimes can work as indicators of regime shifts in the levels of
protein concentration.  However, these indicators can also produce
false alarms due to statistical limitations.  We also performed
sensitivity analysis for the best choice of statistical parameters to
be used in time series analysis as to avoid false alarms.  When the
variance and autocorrelation fails to predict regime shifts, we have
shown that other indicator like conditional heteroskedasticity can be
successful.  The implication of EWS as bio-markers for complex diseases
demands experimental verification and is a future challenge for
experimental biologists.  For predicting regime shifts in gene
expression in experiments, one can use single cell flow cytometry
measurements which gives rapid analysis of multiple characteristics of
single cell \cite{Is2003,BeHa:2009}.  Flow cytometry monitors the
distribution of number of proteins in a cell culture.

Further work on extending the kind of analysis presented here to more
complex gene networks is needed.  Earlier studies in the direction of
understanding and predicting regime shifts in gene expression advanced
our perception, however there is still lack of quantitative
understanding of regime shifts in genetic networks due to its inherent
complexity.  The main advantage of the mathematical formalism adopted
in this paper is that it is simple and easy to understand.  We hope
that, this reductionist approach could form the basis for more
rigorous studies on regime shifts of complex gene networks.  Moreover,
in this study like many other studies on gene expression we have
employed the Langevin and Fokker-Planck description due to their
simplicity and analytic tractability \cite{Ga85,Ha00,Gh12}, but one
can also use the master equation and its Monte-Carlo simulation
\cite{Ga85}.  Finally, acquiring in depth knowledge about the factors
those drive shifts in gene expression states could have significant
impact in clinical biology.

\begin{acknowledgments}
P.S.D. acknowledges financial support from the SERB, Department of
Science and Technology (DST), Govt. of India [Grant No.:
  YSS/2014/000057].
\end{acknowledgments}


\begin{thebibliography}{10}

\bibitem{Book_Sch}
M.~Scheffer.
\newblock {\em Critical transitions in nature and society}.
\newblock Princeton University Press, 2009.

\bibitem{ScJo09}
M.~Scheffer, J.~Bascompte, W.~A. Brock, V.~Brovkin, S.~R. Carpenter, V.~Dakos,
  H.~Held, E.~H van Nes, M.~Rietkerk, and G.~Sugihara.
\newblock {Early-warning signals for critical transitions}.
\newblock {\em Nature}, 461:53--59, 2009.

\bibitem{Sc01}
M.~Scheffer, S.~R. Carpenter, J.~A. Foley, C.~Folke, and B.~Walker.
\newblock Catastrophic shifts in ecosystems.
\newblock {\em Nature}, 413:591--596, 2001.

\bibitem{ScCa2003}
M.~Scheffer and S.~R. Carpenter.
\newblock Catastrophic regime shifts in ecosystems: linking theory to
  observation.
\newblock {\em Trends in Ecology \& Evolution}, 18(12):648--656, 2003.

\bibitem{Wang:2012}
R.~Wang, J.~A. Dearing, P.~G. Langdon, E.~Zhang, X.~Yang, V.~Dakos, and
  M.~Scheffer.
\newblock {Flickering gives early warning signals of a critical transition to a
  eutrophic lake state}.
\newblock {\em Nature}, 492:419--422, 2012.

\bibitem{LeHeKr08}
T.~M. Lenton, H.~Held, E.~Kriegler, J.~W. Hall, W.~Lucht, S.~Rahmstorf, and
  H.~J. Schellnhuber.
\newblock Tipping elements in the earth's climate system.
\newblock {\em Proceedings of the National Academy of Sciences USA},
  105(6):1786--1793, 2008.

\bibitem{Ve05}
J.~G. Venegas, T.~Winkler, G.~Musch, M.F.V. Melo, D.~Layfield, N.~Tgavalekos,
  A.J. Fischman, R.J. Callahan, G.~Bellani, and G.~Bellani.
\newblock Self-organized patchiness in asthma as a prelude to catastrophic
  shifts.
\newblock {\em Nature}, 434:777--782, 2005.

\bibitem{Mc03}
P.~E. McSharry, L.~A. Smith, and L.~Tarassenko.
\newblock Prediction of epileptic seizures: are nonlinear methods relevant?
\newblock {\em Nature Medicine}, 9:241--242, 2003.

\bibitem{kor2014}
K.~S. Korolev, J.~B. Xavier, and J.~Gore.
\newblock Turning ecology and evolution against cancer.
\newblock {\em Nature Reviews Cancer}, 14(5):371--380, 2014.

\bibitem{Ri2016}
M.~G.~O. Rikkert, V.~Dakos, T.~G. Buchman, R.~de~Boer, L.~Glass, A.~O. Cramer,
  S.~Levin, E.~van Nes, G.~Sugihara, M.~D. Ferrari, and E.~A. Tolner.
\newblock Slowing down of recovery as generic risk marker for acute severity
  transitions in chronic diseases.
\newblock {\em Critical Care Medicine}, 44(3):601--606, 2016.

\bibitem{MaLeSu08}
R.~M. May, S.~A. Levin, and G.~Sugihara.
\newblock Complex systems: Ecology for bankers.
\newblock {\em Nature}, 451:893--895, 2008.

\bibitem{FoWh15}
J.~M. Fox and G.~M. Whitesides.
\newblock Warning signals for eruptive events in spreading fires.
\newblock {\em Proceedings of the National Academy of Sciences USA},
  112(8):2378--2383, 2015.

\bibitem{GoSh:2016}
E.~A. Gopalakrishnan, Y.~Sharma, T.~John, P.~S. Dutta, and R.~I. Sujith.
\newblock Early warning signals for critical transitions in a thermoacoustic
  system.
\newblock {\em Scientific Reports}, 6:35310, 2016.

\bibitem{Scheffer:2012sc}
M.~Scheffer, S.~R. Carpenter, T.~M. Lenton, J.~Bascompte, W.~A. Brock,
  V.~Dakos, J.~van~de Koppel, I.~A. van~de Leemput, S.~A. Levin, E.~H. van Nes,
  M.~Pascual, and J.~Vandermeer.
\newblock {Anticipating critical transitions}.
\newblock {\em Science}, 338:344--348, 2012.

\bibitem{DaCa15}
V.~Dakos, S.~R. Carpenter, E.~H. van Nes, and M.~Scheffer.
\newblock Resilience indicators: prospects and limitations for early warnings
  of regime shifts.
\newblock {\em Philosophical Transactions of the Royal Society B: Biological
  Sciences}, 370:20130263, 2015.

\bibitem{SKP14}
Y.~Sharma, K.~C. Abbott, P.~S. Dutta, and A.~K. Gupta.
\newblock Stochasticity and bistability in insect outbreak dynamics.
\newblock {\em Theoretical Ecology}, 8:163--174, 2015.

\bibitem{Sh2016}
Y.~Sharma, P.~S. Dutta, and A.~K. Gupta.
\newblock Anticipating regime shifts in gene expression: The case of an
  autoactivating positive feedback loop.
\newblock {\em Physical Review E}, 93(3):032404, 2016.

\bibitem{Dakos:2012pone}
V.~Dakos, S.~R. Carpenter, W.~A. Brock, A.~M. Ellison, V.~Guttal, A.~R. Ives,
  S.~K{\'e}fi, V.~Livina, D.~A. Seekell, E.~H. van Nes, and M.~Scheffer.
\newblock {Methods for Detecting Early Warnings of Critical Transitions in Time
  Series Illustrated Using Simulated Ecological Data}.
\newblock {\em PLoS One}, 7:e41010, 2012.

\bibitem{CaBr08}
S.~R. Carpenter, W.~A. Brock, J.~J. Cole, J.~F. Kitchell, and M.~L. Pace.
\newblock {Leading indicators of trophic cascades}.
\newblock {\em Ecology Letters}, 11:128--138, 2008.

\bibitem{GuJa08}
V.~Guttal and C.~Jayaprakash.
\newblock {Changing skewness: an early warning signal of regime shifts in
  ecosystems}.
\newblock {\em Ecology Letters}, 11:450--460, 2008.

\bibitem{sc2015}
M.~Scheffer, S.~R. Carpenter, V.~Dakos, and E.~H. van Nes.
\newblock Generic indicators of ecological resilience: inferring the chance of
  a critical transition.
\newblock {\em Annual Review of Ecology, Evolution, and Systematics},
  46:145--167, 2015.

\bibitem{Se11}
D.~A. Seekell, S.~R. Carpenter, and M.~L. Pace.
\newblock {Conditional heteroscedasticity as a leading indicator of ecological
  regime shifts}.
\newblock {\em The American Naturalist}, 178:442--451, 2011.

\bibitem{En1982}
R.~F. Engle.
\newblock Autoregressive conditional heteroscedasticity with estimates of the
  variance of united kingdom inflation.
\newblock {\em Econometrica: Journal of the Econometric Society},
  50(4):987--1007, 1982.

\bibitem{PaPaBo13}
M.~Pal, A.~K. Pal, S.~Ghosh, and I.~Bose.
\newblock Early signatures of regime shifts in gene expression dynamics.
\newblock {\em Physical Biology}, 10(3):036010, 2013.

\bibitem{gl15}
L.~Glass.
\newblock Dynamical disease: Challenges for nonlinear dynamics and medicine.
\newblock {\em Chaos: An Interdisciplinary Journal of Nonlinear Science},
  25:097603, 2015.

\bibitem{TrAn15}
C.~Trefois, P.~M. Antony, J.~Goncalves, A.~Skupin, and R.~Balling.
\newblock Critical transitions in chronic disease: transferring concepts from
  ecology to systems medicine.
\newblock {\em Current Opinion in Biotechnology}, 34:48--55, 2015.

\bibitem{ChLi12}
L.~Chen, R.~Liu, Z-P. Liu, M.~Li, and K.~Aihara.
\newblock Detecting early-warning signals for sudden deterioration of complex
  diseases by dynamical network biomarkers.
\newblock {\em Scientific Reports}, 2:342, 2012.

\bibitem{Van2014}
I.~A. van~de Leemput, M.~Wichers, A.~O. Cramer, D.~Borsboom, F.~Tuerlinckx,
  P.~Kuppens, E.~H. van Nes, W.~Viechtbauer, E.~J. Giltay, S.~H. Aggen, and
  C.~Derom.
\newblock Critical slowing down as early warning for the onset and termination
  of depression.
\newblock {\em Proceedings of the National Academy of Sciences USA},
  111(1):87--92, 2014.

\bibitem{La2012}
J.~Lagro, N.~C. Laurenssen, B.~W. Schalk, Y.~Schoon, J.~A. Claassen, and
  M.~G.~O. Rikkert.
\newblock Diastolic blood pressure drop after standing as a clinical sign for
  increased mortality in older falls clinic patients.
\newblock {\em Journal of Hypertension}, 30(6):1195--1202, 2012.

\bibitem{kra2012}
M.~A. Kramer, W.~Truccolo, U.~T. Eden, K.~Q. Lepage, L.~R. Hochberg, E.~N.
  Eskandar, J.~R. Madsen, J.~W. Lee, A.~Maheshwari, E.~Halgren, C.~J. Chu, and
  S.~S. Cash.
\newblock Human seizures self-terminate across spatial scales via a critical
  transition.
\newblock {\em Proceedings of the National Academy of Sciences USA},
  109(51):21116--21121, 2012.

\bibitem{Sc2013}
M.~Scheffer, A.~van~den Berg, and M.~D. Ferrari.
\newblock Migraine strikes as neuronal excitability reaches a tipping point.
\newblock {\em PLoS One}, 8(8):e72514, 2013.

\bibitem{Sm98}
P.~Smolen, D.~A. Baxter, and J.~H. Byrne.
\newblock Frequency selectivity, multistability, and oscillations emerge from
  models of genetic regulatory systems.
\newblock {\em American journal of Physiology}, 274:C531--C542, 1998.

\bibitem{KaEl05}
M.~Kaern, T.~C. Elston, W.~J. Blake, and J.~J. Collins.
\newblock Stochasticity in gene expression: from theories to phenotypes.
\newblock {\em Nature Reviews Genetics}, 6:451--464, 2005.

\bibitem{Ha00}
J.~Hasty, J.~Pradines, M.~Dolnik, and J.~J. Collins.
\newblock Noise-based switches and amplifiers for gene expression.
\newblock {\em Proceedings of the National Academy of Sciences USA},
  97:2075--2080, 2000.

\bibitem{LiJi04}
Q.~Liu and Y.~Jia.
\newblock Fluctuations-induced switch in the gene transcriptional regulatory
  system.
\newblock {\em Physical Review E}, 70:041907, 2004.

\bibitem{Ch08}
{Z. Cheng}, {F. Liu}, {X. Zhang}, and {W. Wang}.
\newblock {R}obustness analysis of celular memory in an autoactivating positive
  feedback system.
\newblock {\em FEBS Letters}, 582:3776–--3782, 2008.

\bibitem{Frcasaib12}
{D. Frigola}, {L. Casanellas}, {J. M. Sancho}, and {M. Ibañes}.
\newblock {A}symmetric stochastic switching driven by intrinsic molecular
  noise.
\newblock {\em PLoS One}, 7:e31407, 2012.

\bibitem{Gh12}
S.~Ghosh, S.~Banerjee, and I.~Bose.
\newblock Emergent bistability: Effects of additive and multiplicative noise.
\newblock {\em The European Physical Journal E}, 35(2):1--14, 2012.

\bibitem{Ro2005}
N.~Rosenfeld, J.~W. Young, U.~Alon, P.~S. Swain, and M.~B. Elowitz.
\newblock Gene regulation at the single-cell level.
\newblock {\em Science}, 307(5717):1962--1965, 2005.

\bibitem{Si2006}
A.~Sigal, R.~Milo, A.~Cohen, N.~Geva-Zatorsky, Y.~Klein, Y.~Liron,
  N.~Rosenfeld, T.~Danon, N.~Perzov, and U.~Alon.
\newblock Variability and memory of protein levels in human cells.
\newblock {\em Nature}, 444(7119):643--646, 2006.

\bibitem{ShOSw08}
V.~Shahrezaei, J.~F. Ollivier, and P.~S. Swain.
\newblock Colored extrinsic fluctuations and stochastic gene expression.
\newblock {\em Molecular Systems Biology}, 4(1):196, 2008.

\bibitem{du2008}
M.~J. Dunlop, R.~S. Cox, J.~H. Levine, R.~M. Murray, and M.~B. Elowitz.
\newblock Regulatory activity revealed by dynamic correlations in gene
  expression noise.
\newblock {\em Nature Genetics}, 40(12):1493--1498, 2008.

\bibitem{Dunlop:thesis2008}
M.~J. Dunlop.
\newblock {\em Dynamics and Correlated Noise in Gene Regulation}.
\newblock PhD thesis, California Institute of Technology, Pasadena, California,
  2008.

\bibitem{Kl1988}
M.~M. Klosek-Dygas, B.~J. Matkowsky, and Z.~Schuss.
\newblock Colored noise in dynamical systems.
\newblock {\em SIAM Journal on Applied Mathematics}, 48(2):425--441, 1988.

\bibitem{BeSeSe01}
{A. Becksei}, {B. Seraphin}, and {L. Serrano}.
\newblock {P}ositive feedback in eukaryotic gene networks: cell differentiation
  by graded to binary response.
\newblock {\em The EMBO}, 20:2528–--2535, 2001.

\bibitem{TyChNo03}
{J. J. Tyson}, {K. C. Chen}, and {B. Novak}.
\newblock {S}niffers, buzzers, toggles and blinkers: dynamics of regulatory and
  signaling pathways in the cell.
\newblock {\em Current Opinion in Cell Biology}, 15:221–--231, 2003.

\bibitem{Is2003}
F.~J Isaacs, J.~Hasty, C.~R. Cantor, and J.~J. Collins.
\newblock Prediction and measurement of an autoregulatory genetic module.
\newblock {\em Proceedings of the National Academy of Sciences USA},
  100(13):7714--7719, 2003.

\bibitem{Tha2001}
M.~Thattai and A.~Van~Oudenaarden.
\newblock Intrinsic noise in gene regulatory networks.
\newblock {\em Proceedings of the National Academy of Sciences USA},
  98(15):8614--8619, 2001.

\bibitem{ZhYaTa11}
{X. Zheng and X. Yang and Y. Tao}.
\newblock Bistability, probability transition rate and first-passage time in an
  autoactivating positive-feedback loop.
\newblock {\em PLoS One}, 6:e17104, 2011.

\bibitem{Ra03}
C.~V. Rao and A.P. Arkin.
\newblock Stochastic chemical kinetics and the quasi-steady-state assumption:
  application to the gillespie algorithm.
\newblock {\em The Journal of Chemical Physics}, 118(11):4999--5010, 2003.

\bibitem{WeBu13}
{M. Weber and J. Buceta}.
\newblock Stochastic stabilization of phenotypic states: The genetic bistable
  switch as a case study.
\newblock {\em PLoS One}, 7:e73487, 2013.

\bibitem{Ga85}
C.~W. Gardiner.
\newblock {\em Handbook of Stochastic Methods: {F}or Physics, Chemistry and the
  Natural Sciences}.
\newblock Springer--Verlag, Berlin, 2nd edition, 1985.

\bibitem{Sa1982}
J.~M. Sancho, M.~San~Miguel, S.~L. Katz, and J.~D. Gunton.
\newblock Analytical and numerical studies of multiplicative noise.
\newblock {\em Physical Review A}, 26(3):1589, 1982.

\bibitem{liang2004}
G.~Y. Liang, L.~Cao, and D.~J. Wu.
\newblock Approximate fokker--planck equation of system driven by
  multiplicative colored noises with colored cross-correlation.
\newblock {\em Physica A: Statistical Mechanics and its Applications},
  335(3):371--384, 2004.

\bibitem{WhoRle84}
{W. Horsthemke} and {R. Lefever}.
\newblock {\em Noise-Induced Transitions}.
\newblock Springer, Berlin, 1984.

\bibitem{Ro2010}
S.~Rothman.
\newblock How is the balance between protein synthesis and degradation
  achieved?
\newblock {\em Theoretical Biology and Medical Modelling}, 7(1):1, 2010.

\bibitem{Cui2011}
J.~Cui, Y.~Chen, W.C. Chou, L.~Sun, L.~Chen, J.~Suo, Z.~Ni, M.~Zhang, X.~Kong,
  L.L. Hoffman, and J.~Kang.
\newblock An integrated transcriptomic and computational analysis for biomarker
  identification in gastric cancer.
\newblock {\em Nucleic acids research}, 39(4):1197--1207, 2011.

\bibitem{Jin2009}
G.~Jin, X.~Zhou, K.~Cui, X.S Zhang, L.~Chen, and S.T.C Wong.
\newblock Cross-platform method for identifying candidate network biomarkers
  for prostate cancer.
\newblock {\em IET Systems Biology}, 3(6):505--512, 2009.

\bibitem{DH01}
D.~J. Higham.
\newblock An algorithmic introduction to numerical simulation of stochastic
  differential equations.
\newblock {\em SIAM Review}, 43(3):525--546, 2001.

\bibitem{Drake:2013}
J.~M Drake.
\newblock Early warning signals of stochastic switching.
\newblock {\em Proceedings of the Royal Society of London B: Biological
  Sciences}, 280(1766):20130686, 2013.

\bibitem{Boet:2013}
C.~Boettiger and A.~Hastings.
\newblock No early warning signals for stochastic transitions: insights from
  large deviation theory.
\newblock {\em Proceedings of the Royal Society of London B: Biological
  Sciences}, 280(1766):20131372, 2013.

\bibitem{Ak1998}
H.~Akaike.
\newblock Information theory and an extension of the maximum likelihood
  principle.
\newblock In {\em Selected Papers of Hirotugu Akaike}, pages 199--213.
  Springer, 1998.

\bibitem{BeHa:2009}
M.~Bennett and J.~Hasty.
\newblock Microfluidic devices for measuring gene network dynamics in single
  cells.
\newblock {\em Nature Reviews Genetics}, 10:628--638, 2009.

\end{thebibliography}

\end{document}